\documentclass[english,journal=jctcce,manuscript=article,email=true,etalmode=truncate,maxauthors=0,doi=true]{achemso}

\usepackage{graphicx}
\usepackage{siunitx}
\usepackage{comment}
\graphicspath{{figures/}}
\usepackage{physics}
\usepackage{multirow}
\usepackage{xcolor}
\usepackage{caption}
\usepackage{subcaption}
\usepackage{xcolor}
\usepackage{bm}
\usepackage{eufrak} 
\usepackage{yfonts} 
\usepackage[title,titletoc]{appendix}
\usepackage{diagbox}
\usepackage{booktabs}
\usepackage{adjustbox}

\usepackage{hyperref} 
\hypersetup{colorlinks=true, citecolor=blue, urlcolor=blue, linkcolor=blue}
\usepackage{cleveref}
  	\crefname{figure}{Figure}{Figures}
  	\crefname{table}{Table}{Tables}
  	\crefname{equation}{Eq.}{Eqs.}
  	\crefname{section}{Section}{Sections}
  	\crefname{subsection}{Section}{Sections}
  	\crefname{subsubsection}{Section}{Sections}
  	\crefname{algorithm}{Algorithm}{Algorithms}

\newcommand{\code}[1]{\texttt{#1}}
\newcommand\vartextvisiblespace[1][.5em]{%
  \makebox[#1]{%
    \kern.07em
    \vrule height.3ex
    \hrulefill
    \vrule height.3ex
    \kern.07em
  }
}
\usepackage{todonotes}

\title{Towards Balanced Description of Ground and Excited States with Transcorrelated F12 Methods} 

\author{Conner Masteran}
\author{Bimal Gaudel}
\author{Edward F. Valeev}
\email{efv@vt.edu}
\affiliation{Department of Chemistry, Virginia Tech, Blacksburg, VA 24061}

\begin{document}

\date{\today}

\begin{abstract}
By correlating only the 1-particle states occupied in the reference determinant the conventional design for the single-reference R12/F12 explicitly-correlated methods biases them towards the ground state description thereby making treatment of response properties of the ground state, and energies and other properties of excited states less robust. While the use of multireference methods and/or extensions of the standard SP projected geminals can achieve a more balanced description of ground and excited states, here we show that the same goals can be achieved by extending the action of F12 correlators to the occupied and valence unoccupied 1-particle states only. This design choice reflects the strong dependence of the optimal correlation lengthscale of the F12 ansatz on the orbital energies/structure, and helps to avoid the unphysical raising of the ground state energy if the F12 geminals are used to correlate pairs of all 1-particle states. The improved F12 geminal design is incorporated into the unitary transcorrelation framework to produce a unitary 2-body Hamiltonian that incorporates the short-range dynamical correlation physics for ground and low-energy excited states in a balanced manner. This explicitly-correlated effective Hamiltonian reduces the basis set requirement on the correlation-consistent basis cardinal number by 1 or more over the uncorrelated counterpart for the description of the ground state coupled-cluster singles and doubles (CCSD) energies, the vertical excitation energies and harmonic vibrational frequencies of equation-of-motion CCSD low-energy excited states.
\end{abstract}

\section{Introduction}\label{sec:intro}

The utility of interelectronic distances as explicit variables in the definition of many-electron wave functions has been known since the work of Hylleraas\cite{VRG:hylleraas:1929:ZFP} and Slater\cite{VRG:slater:1928:PRa,VRG:slater:1928:PRb}. The explicit use of such coordinates in the {\em explicitly-correlated} methods greatly speeds up the basis set convergence by modeling more efficiently the known analytic behavior of the wave function near the electron-electron cusps of the exact wave function\cite{VRG:kato:1957:CPAM,VRG:pack:1966:JCP}.
There are several groups of explicitly-correlated approaches in use today:
\begin{itemize}
\item {\bf high-precision methods}: involve unfactorizable many-particle integrals with most or all particles at once that are evaluated analytically; examples include Hylleraas-CI,\cite{VRG:ruiz:2021:AiQC} explicitly-correlated Gaussian geminals\cite{VRG:mitroy:2013:RMP}, and free iterative complement interaction (ICI) approach of Nakatsuji\cite{VRG:nakatsuji:2004:PRL}.
\item {\bf configuration-space Jastrow factor}: 
methods involving unfactorizable many-particle integrals with most or all particles at once that are evaluated stochastically or avoided by serving as a trial wave function; examples include single- and multiconfiguration Slater-Jastrow wave functions in variational and diffusion quantum Monte-Carlo methods.\cite{VRG:foulkes:2001:RMP}
\item {\bf configuration-space transcorrelation}: involves unfactorizable integrals with at most 3 particles that are evaluated analytically;\cite{VRG:boys:1969:PRSLA,VRG:boys:1969:PRSMPES}
\item {\bf R12/F12 methods}: involve unfactorizable integrals with up to 4 particles that are evaluated analytically (up to 2-particle integrals) or by the resolution-of-the-identity (3- and 4-particle integrals).\cite{VRG:kutzelnigg:1985:TCA,VRG:kong:2012:CR,VRG:hattig:2012:CR,VRG:ten-no:2012:WIRCMS,VRG:ten-no:2012:TCA}
\end{itemize}
By only requiring the exact evaluation of (nonstandard) 2-particle integrals the F12 methods are most practical among all explicitly-correlated methods, available routinely in several packages and with demonstrated applications to hundreds and thousands of atoms.\cite{VRG:pavosevic:2016:JCP,VRG:ma:2018:JCTC,VRG:kumar:2020:JCP,VRG:wang:2023:JCTC}
    Another key advantage of modern F12 methods over most other explicitly-correlated methods is that no problem-specific nonlinear optimization of the parameters of the explicitly-correlated terms is required. This is due to the use of the explicitly-correlated terms (``geminals'') to handle {\em only} the largely universal physics\cite{VRG:kato:1957:CPAM,VRG:pack:1966:JCP} of the short-range electron correlation near the electron-electron cusp; the conventional superpositions of orbital products (Slater determinants) account for the rest of nonuniversal physics of longer-range correlations (such as dispersion interactions). Lastly, the spin dependence of the cusp conditions\cite{VRG:pack:1966:JCP} can be rigorously satisfied in the modern F12 framework\cite{VRG:ten-no:2004:JCP} unlike in the purely configuration-space methods\cite{VRG:huang:1998:JCP}.

The vast majority of development and applications of the F12 framework have focused on the properties of ground states, such as thermochemistry and physics of noncovalent interactions, which are particularly impacted by the basis set errors from short-range electron correlations. Most of the time the lowest vertical excitation energies, on the other hand, often exhibit much smaller basis set errors. This is due to the robust cancellation of dynamical correlation errors in the ground and excited states. Nevertheless, the basis set errors can be substantial (especially for states with a Rydberg character\cite{VRG:shiozaki:2010:JCPa}) and difficult to suppress. F12 methods are also expected to be important for an accurate description of excited state potential energy surfaces. Hence, it is highly desirable to deploy the F12 framework for the description of excited states and response properties. And, although multiple electronic states can be treated on equal footing by the F12 extensions of multi-configuration wave function methods\cite{VRG:shiozaki:2010:JCPa,VRG:shiozaki:2011:JCPa}, it is highly desirable to deploy the F12 framework in the context of conventional single-reference approaches like the coupled-cluster theory that can often model electron correlation of ground and low-energy excited states with unrivaled accuracy.
A hindrance to such developments is that F12 methods recover more correlation energy for the ground than excited states leading to an imbalanced description between the states. This imbalance leads to artificially high vertical excitation energies, as was first observed by Fliegl et al.\cite{cmast:Fliegl:2006:JCP} This imbalance in the description of ground vs. excited states can be partially attributed to the construction of the geminal terms. The original orbital-invariant ansatz of the F12 theory\cite{VRG:klopper:1991:CPL} and the de facto standard ``SP'' ansatz\cite{VRG:ten-no:2004:JCP} only explicitly correlate pairs of occupied (or {\em hole}, in the quasiparticle picture) orbitals. There have been several approaches that use geminals to also correlate particles in unoccupied (virtual, or {\em particle}) orbitals.
Neiss, H{\"a}ttig, and Klopper extended the set of geminal-generating orbitals to include a subset of particle orbitals selected according to the occupation numbers of MP2 densities\cite{cmast:Neiss:2006:JCP,VRG:neiss:2007:JCP}.
K\"{o}hn extended the SP ansatz to include geminal excitations from hole-particle orbital pairs.\cite{cmast:Kohn:2009:JCP}  K\"{o}hn's extended SP (XSP) ansatz has been used for description of response properties in the linear response coupled-cluster framework \cite{cmast:Kohn:2009:JCP,cmast:Kohn:2019:JCP} and for prediction of excitation energies in the context of explicitly-correlated EOM-CC methods by Bokhan and Ten-no\cite{cmast:Ten-no:2010:JCP} and later in the STEOM form by Bokhan et al.\cite{cmast:Bartlett:2015:JCP}
Geminal-generating orbitals also have to include unoccupied orbitals in the context of energy-dependent F12 single-particle propagator methods.\cite{VRG:pavosevic:2017:JCP,VRG:teke:2019:JCP}

Although the use of geminals for correlating particles in the reference-unoccupied orbitals does improve the description of excitation energies and related quantities like response functions and self-energies, the price seems to be a {\em worse} description of the ground state compared to the SP ansatz.\cite{cmast:Kohn:2009:JCP} The XSP ansatz can address this issue by introducing additional parameters, such as in the XSP$_\mathrm{opt}$ approach\cite{cmast:Kohn:2009:JCP}, but unfortunately the numerical optimization of geminal-related amplitudes (practiced long before the introduction of the SP ansatz but largely abandoned since the SP ansatz introduction) can be ill-posed. In this paper we reexamine the problem of constructing the F12 ansatz appropriate for the low-lying excited states {\em without} sacrificing the accuracy of the SP ansatz for the ground states {\em and without} the need to introduce additional parameters (the latter is particularly important in the context of transcorrelated form of F12 methods used here). Our key hypothesis is that the single-parameter exponential correlation factors used in the modern F12 toolkit are tuned primarily for valence orbitals and are less appropriate for description of orbital pairs that include nonvalence orbitals. The idea that geminal definition must be adjusted for nonvalence orbitals is already well-known from studies of correlation energies including core electrons; e.g., Werner et al. found that the optimal correlation factor exponents differ substantially between core-core(cc), core-valence(cv), and valence-valence(vv) orbital pairs\cite{cmast:Manby:2011:MolPhys}. In general, the optimal exponents were the largest for cc contributions and the smallest for vv contributions.

Our investigation of the geminal basis construction will use the {\em a priori} form of the F12 technology, where the Hamiltonian is transformed using the F12 wave operator first. A particular form of F12 transcorrelation used here is based on the approximate unitary (canonical) transcorrelated approach of Yanai and Shiozaki\cite{VRG:yanai:2012:JCP} who referred to the approach as CT-F12 and used by ourselves\cite{VRG:motta:2020:PCCP,cmast:Kumar:2022:JCTC} and others\cite{VRG:kersten:2016:JCP} to significantly reduce the basis set errors. The errors in the original formalism and its implementation were recently corrected using an automated implementation.\cite{cmast:Masteran:2023:JCP} The unitary F12 TC formalism was shown to be as robust as traditional F12 approaches for ground states\cite{cmast:Masteran:2023:JCP}. Unlike the configuration-space transcorrelation of Boys and Handy\cite{VRG:boys:1969:PRSLA,VRG:boys:1969:PRSMPES,VRG:handy:1969:JCP,VRG:handy:1971:MP} which is the subject of intense recent attention,\cite{cmast:Schraivogel:2021,VRG:baiardi:2022:JCTC,VRG:kats:2024:FD,VRG:szenes:2024:FD,VRG:lee:2023:JCTC} the F12-style transcorrelation avoids nonfactorizable 3-particle integrals (via approximate resolution of the identity), nonlinear optimization, or spin-contamination. The unitary F12 transcorrelation also has some advantages over the nonunitary counterpart\cite{VRG:ten-no:2023:JCP} in that the transformed (downfolded) Hamiltonian is Hermitian and only includes 2-particle interactions, hence it can be used with most electronic structure models without modification. Here we used it in the context of excited-state single-reference coupled-cluster models.

The rest of the paper is structured as follows. In \cref{sec:formalism} we describe the unitary F12 TC formalism and introduce several new F12 geminal ansatze. \cref{sec:methods} contains the pertinent technical details of our computational experiments. In \cref{sec:results} we report on the performance of the standard and new F12 geminal ansatze for ground- and excited-state energies and properties. \cref{sec:summary} contains a summary of our findings.

\section{Formalism}\label{sec:formalism}

\subsection{Hermitian F12 Transcorrelation}

\begin{figure*}[ht]
\includegraphics[width=0.90\textwidth]{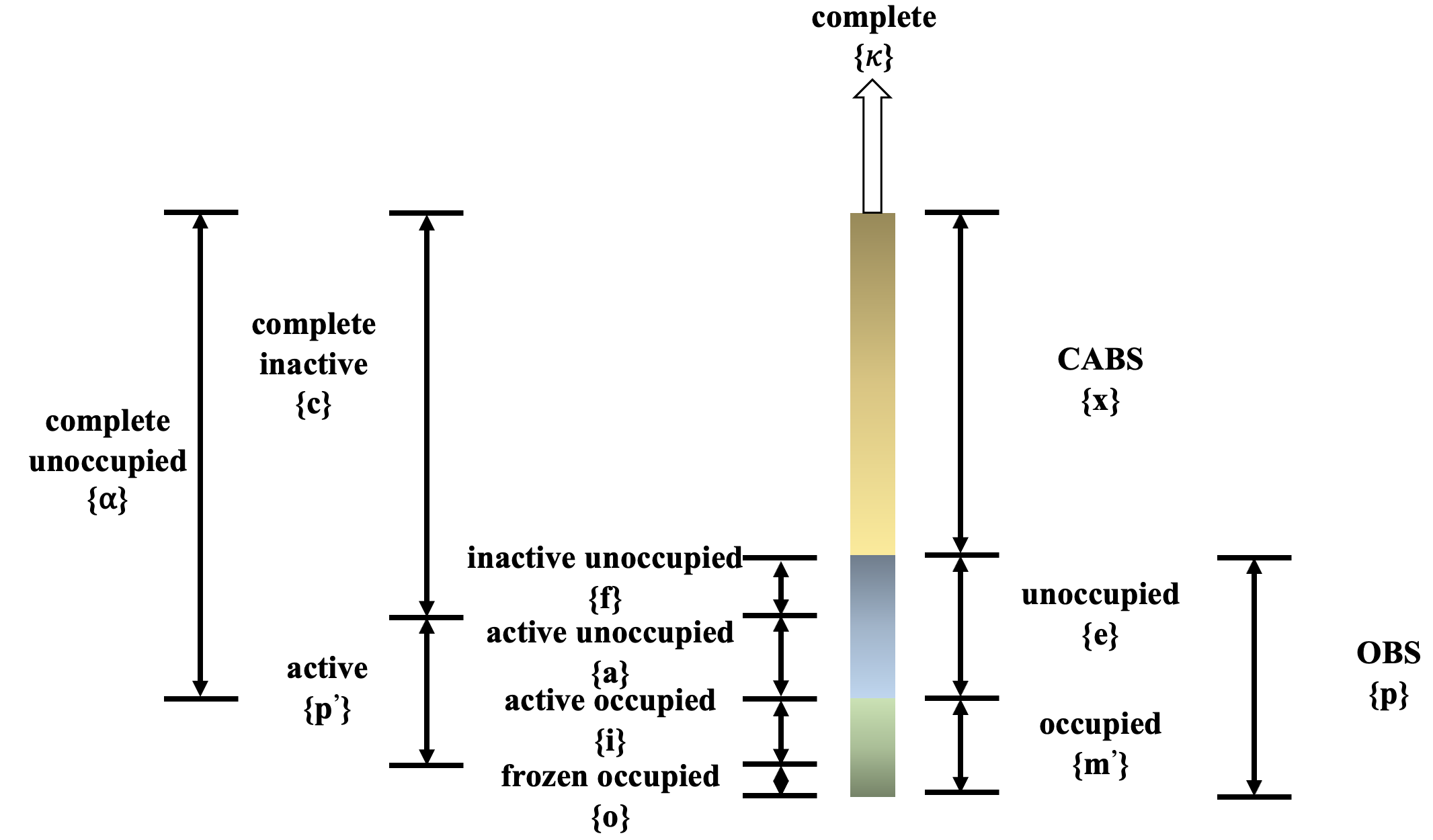}
    \caption{Schematic diagram of the single-particle state spaces used in this text.}
    \label{figure:spaces}
\end{figure*}

Transcorrelated Hamiltonians are generated by a similarity transformation with the operator $e^{\hat{A}}$, where $\hat{A}$ is an operator that includes dependence on the interparticle distances and designed such that the TC Hamiltonian,
\begin{equation}
\hat{\bar{H}} = e^{-\hat{A}} \hat{H} e^{\hat{A}} \quad ,
\label{eq:H-TC}
\end{equation}
is free of 2-electron singularities.
In the original transcorrelation method of Boys and Handy\cite{VRG:boys:1969:PRSLA,VRG:boys:1969:PRSMPES,VRG:handy:1969:JCP} and its subsequent reincarnations\cite{cmast:Hino:2001,cmast:Hino:2002,cmast:Cohen:2019,cmast:Schraivogel:2021} $\hat{A}$ is a multiplicative configuration-space operator which can assume very general forms,\cite{VRG:nooijen:1998:JCP} with the resulting {\em non-Hermitian} TC Hamiltonian having a closed analytic form usually including up to 3-particle terms.
In contrast, projective transcorrelation methods
use a Fock-space correlator $\hat{A}$ constructed in the spirit of R12/F12 methods; the resulting TC Hamiltonians do not have an analytic form but can be evaluated efficiently using the robust approximations of the R12/F12 methods. The original projective transcorrelation approach of Yanai-Shiozaki produced a {\em Hermitian} TC Hamiltonian as a nontruncating Baker-Campbell-Hausdorff (BCH) series
by using an {\em antihermitian} $\hat{A}$ and ad hoc truncating the BCH series to ensure cancellation of Coulomb electron-electron singularities to first order; it is not a coincidence that the resulting complexity of the many-particle integrals (and thus the short-range correlation physics) is similar to that of the MP2-F12 and approximate CCSD-F12 methods.\cite{VRG:valeev:2008:PCCPP,VRG:adler:2007:JCP,VRG:hattig:2010:JCP}
Although the recently-proposed {\em non-Hermitian} projective transcorrelation $\hat{A}$ can formally assure rigorous truncation of the BCH series for $\hat{\bar{H}}$ at high particle ranks by using a general excitation-only Fock-space $\hat{A}$,
in practice any projective transcorrelation method must truncate the BCH series to avoid a runaway growth of the particle rank of TC Hamiltonian. Thus, the Hermitian vs. non-Hermitian design pivot is unlikely to grant a definitive advantage.

Thus, here we will focus on the Hermitian projective TC approach that was originally investigated\cite{VRG:yanai:2012:JCP,cmast:Masteran:2023:JCP} with $\hat{A}$ defined to introduce correlations only between pairs of {\em active} reference-occupied single-particle ({\em sp}) states $\{i\}$ (see \cref{figure:spaces} for the sp space glossary):
\begin{align}
    \hat{A}_{\text{II}} = \frac{1}{2} (G_{\alpha_1\alpha_2}^{i_1 i_2}\hat{E}^{\alpha_1\alpha_2}_{i_1i_2} - {G^{\dag}}_{i_1i_2}^{\alpha_1\alpha_2}\hat{E}^{i_1i_2}_{\alpha_1\alpha_2}) \equiv \frac{1}{2} (G_{\alpha_1\alpha_2}^{i_1 i_2}\hat{E}^{\alpha_1\alpha_2}_{i_1i_2} - h.c.) ,
    \label{eq:AII}
\end{align}
where $h.c$ denotes the Hermitian conjugate of the preceding terms.
Here and elsewhere, the Einstein summation convention of duplicate indices is assumed, unless noted explicitly.
In \cref{eq:AII} $\hat{E}$ is the {\em spin-free} fermionic 2-particle transition operator normal-ordered with respect to the genuine (0-particle) vacuum:
\begin{align}
    \hat{E}_{\kappa_1 \kappa_2}^{\kappa_3 \kappa_4} = & \sum_{\sigma \tau} \hat{a}_{\kappa_{3} 
    \sigma}^{\dag}\hat{a}^{\dag}_{\kappa_4 \tau}\hat{a}_{\kappa_2 \tau } \hat{a}_{\kappa_1 \sigma}
    \label{eq:E2}
\end{align}
with $\sigma$ and $\tau$ enumerating the spin state basis, and $G$ and $G^\dagger$ are the matrix elements of the standard SP ansatz projected geminal\cite{VRG:ten-no:2004:JCP,VRG:zhang:2012:JCTC}:
\begin{align}
    G^{i_1i_2}_{\alpha_1\alpha_2} = \bra{\alpha_1\alpha_2}\hat{C} \hat{Q}_{12}f(r_{12})\ket{i_1i_2},
    \label{eq:SP} \\
    {G^{\dag}}_{i_1i_2}^{\alpha_1\alpha_2} = \bra{i_1i_2}f(r_{12})\hat{Q}_{12}\hat{C}\ket{\alpha_1\alpha_2},
    \label{eq:SP_t}
\end{align}
with $\hat{C} = (3 + \hat{P}_{12})/8$ adapting the correlator to the S- and P-wave cusp conditions\cite{VRG:pack:1966:JCP} ($\hat{P}_{12}$ is swaps particles 1 and 2).
 The projector $\hat{Q}_{12}$ used in this work,
 \begin{align}
 \hat{Q}_{12} = 1-\hat{V}_1\hat{V}_2,
 \label{eq:Q12}
 \end{align}
 corresponds to the standard ansatz 2 of the R12/F12 theory\cite{VRG:valeev:2004:JCP,VRG:valeev:2004:CPL} (some denote it as ansatz 3\cite{VRG:werner:2006:JCP}).
 whose effect is to exclude the conventional intra-OBS replacements from $\hat{A}$ that are going to be accounted for by the conventional correlation treatment that follows the transcorrelation. That is, \cref{eq:Q12} excludes the contributions of double replacements from the active occupied orbitals to all unoccupied orbitals (that is, $E^{e_1 e_2}_{i_1 i_2}$). The standard single-parameter Slater-type geminal (STG),\cite{VRG:ten-no:2004:JCP} $f(r_{12}) = -\frac{e^{-\gamma r_{12}}}{\gamma}$, is used, with the recommended value\cite{cmast:Peterson:2008:JCP} of the inverse lengthscale $\gamma$ optimized a priori and tabulated for the most common OBS.
 
With this in place  we can define an effective singularity-free Hamiltonian by the correlator it contains:
\begin{align}
    \hat{\bar{H}}_{\text{II}} = e^{-\hat{A}_{\text{II}}}\hat{H}e^{\hat{A}_{\text{II}}}
    \label{eq:H_II}
\end{align}
 Although the BCH expansion of \cref{eq:H-TC} generated by the antihermitian $\hat{A}$ does not truncate, the leading-order cancellation of singularities is ensured by keeping the up to quadratic terms:\cite{VRG:yanai:2012:JCP,cmast:Masteran:2023:JCP} 
 \begin{equation}
     e^{-\hat{A}} \hat{H} e^{\hat{A}} \approx \hat{H} + [\hat{H},\hat{A}]_{12} + \frac{1}{2}[[\hat{F},\hat{A}],\hat{A}]_{12} ,
     \label{eq:BCH}
 \end{equation}
 where $\hat{F}$ is the Fock operator. The single and double commutator terms in \cref{eq:BCH} contain up to 3-particle operators that are further approximated (as denoted by ``$_{12}$'') by rewriting them in normal order with respect to a reference state and neglecting the resulting 3-particle normal-ordered operators and cumulants of the 3-particle reduced density matrix (RDM) of the reference.\cite{cmast:Mukherjee:1997:CPL,cmast:Kutzelnigg:1997:JCP,cmast:Kutzelnigg:2010}
 
 The resulting TC Hamiltonian is Hermitian and contains up to 2-particle terms. As documented in Ref. \citenum{cmast:Masteran:2023:JCP}, the performance of transcorrelation for ground-state CC methods with up to quadruple excitations matches or beats the performance of the conventional CC-F12 methods, by reducing the basis set error of the correlation energy by an equivalent of 2 cardinal numbers.
 
 Unfortunately, the TC Hamiltonian produced with \cref{eq:AII} is not appropriate for the treatment of excited states due to its bias towards the ground state. There are several sources of the bias. The first (and most important) factor is the use of states that are occupied and unoccupied in the ground-state reference in the correlator (\cref{eq:AII}). The second factor is the use of the ground-state Fock operator in \cref{eq:BCH}. The third factor is the use of a zeroth-order ground state reference for defining the cumulant-based screening of the 3-particle terms. And the last factor is the use of reference-unoccupied states in projector $\hat{Q}_{12}$ (\cref{eq:Q12}).
 
 The first step towards eliminating these sources of bias was to generalize $\hat{A}$ to correlate pairs of electrons in any active state $p'$ rather than in any occupied states: 
 \begin{equation}\label{eq:APP}
 \hat{A}_\text{PP} = \frac{1}{2} (G_{\alpha_1\alpha_2}^{p'_1p'_2}\hat{E}^{\alpha_1\alpha_2}_{p'_1p'_2} - h.c).
 \end{equation}
 To distinguish correlators in \cref{eq:AII} and \cref{eq:APP} we will refer to them as II and PP, respectively.
 Note that both are sufficiently general to be usable in both single- and multi-determinantal reference states. Although neither ansatz II nor PP ensure full state universality, we expect ansatz PP to provide a better description of low-lying excited states. Relative performance of ansatze II and PP was recently documented by Kumar et al.\cite{cmast:Kumar:2022} but the use of near-minimal bases did not allow us to generalize the findings. The goal of this manuscript is to rectify this gap.
 
 A natural half-step from II to PP ansatz is the hybrid IP ansatz:
 \begin{equation}
 \hat{A}_{\text{IP}} = \left(G_{\alpha_1 \alpha_2}^{i_1 p'_1} \hat{E}^{\alpha_1 \alpha_2}_{i_1 p'_1} - h.c \right) - \hat{A}_{\text{II}},
 \label{eq:AIP}
 \end{equation}
 where the subtraction of $\hat{A}_{\text{II}}$ is used to avoid double inclusion of the II correlations.
  Such choice of $\hat{A}$ introduces correlations between pairs of electrons in which at least one electron is in an occupied orbital $i$ and the other is in any active orbital $p'$. Although this ansatz seems to be an artificial choice for a correlator, its strong connection to the F12 ansatz used\cite{cmast:Kohn:2009:JCP,cmast:Kohn:2019:JCP} in the context of linear response coupled-cluster F12 methods makes it a useful point of comparison.
 
 In this paper we compare the performance of the three correlator ansatze for ground and excited state properties. The use of automated symbolic tensor algebra framework \code{SeQuant}, which was essential for correct implementation of the II version of the F12 TC Hamiltonian in Ref. \citenum{cmast:Masteran:2023:JCP}, allows straigtforward implementation of the IP and PP generalizations. Although the IP ansatz equations are longer than their II and PP counterparts due to the more complex form of \cref{eq:AIP}, all three ansatze can be implemented with the same 3 ``special'' F12 intermediates V, X, and B, which for the PP ansatz read:
 \begin{align}
     V_{p'_1 p'_2}^{p_1 p_2} = & g^{p_1 p_2}_{\alpha_1 \alpha_2} G^{\alpha_1 \alpha_2}_{p'_1 p'_2}
     \label{eq:V} \\
     X_{p'_1 p'_2}^{p'_3 p'_4} = & G^{p'_3 p'_4}_{\alpha_1 \alpha_2} G^{\alpha_1 \alpha_2}_{p'_1 p'_2} 
     \label{eq:X} \\
    B_{p'_1 p'_2}^{p'_3 p'_4} = & G_{p'_1 p'_2}^{\alpha_1 \alpha_2}f^{\alpha_3}_{\alpha_1} G^{p'_3 p'_4}_{\alpha_3 \alpha_2} + G_{p'_1 p'_2}^{\alpha_1 \alpha_2}f^{\alpha_3}_{\alpha_2} G^{p'_3 p'_4}_{\alpha_1 \alpha_3} .
    \label{eq:B}
\end{align}
Generalization of \cref{eq:V,eq:X,eq:B} to the IP case is straightforward. Intermediates V and X are evaluated using the CABS+ approach\cite{VRG:valeev:2004:CPL} and intermediate B is evaluated using the so-called approximation C.\cite{VRG:kedzuch:2005:IJQC}

\subsection{Valence-Constrained Geminals}

 \begin{figure*}
     \centering
         \includegraphics[width=0.90\textwidth]{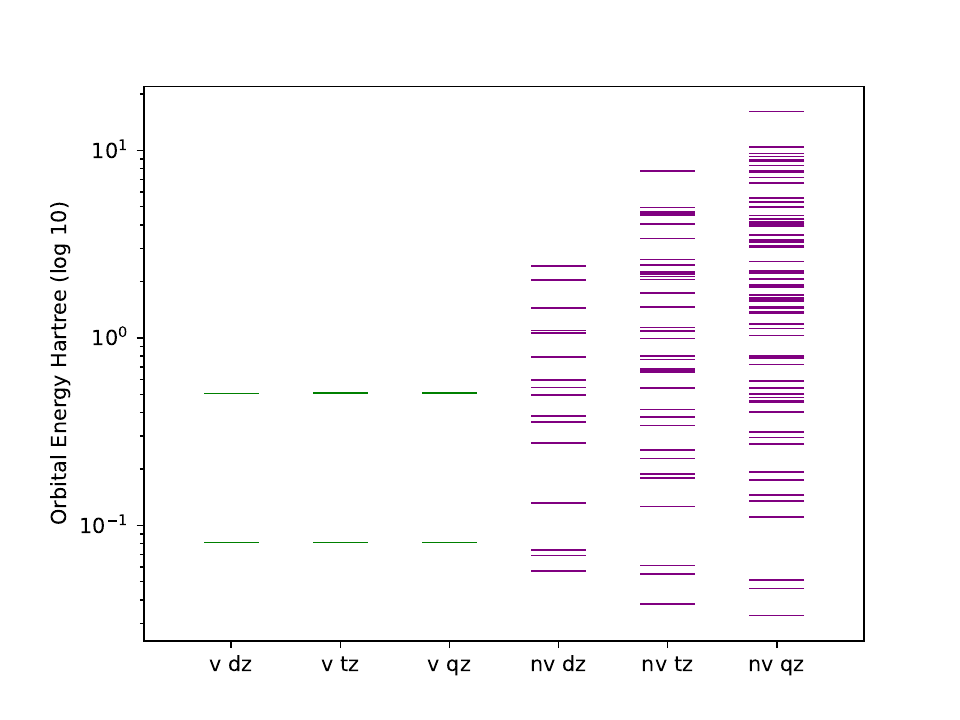}
    \caption{Pseudo-eigenvalues of the unoccupied orbitals of BH in aug-cc-pVXZ basis sets. The valence orbitals (v) are shown on the left in green while the remaining non-valence (nv) unoccupied orbitals are seen on the right in purple. Notice that the size of the active space is independent of the chosen orbital basis set.}
    \label{fig:VVO_figure}
\end{figure*}

 Unfortunately, the use of IP and PP ansatze produces ground state energies that are consistently higher than those from the more restrictive II ansatz. There is some precedent in the literature for such raising of the ground state correlation energy by straightforward use of geminals for correlation of virtual orbitals. Namely, Figures 2 and 3 in Andreas K\"ohn's paper\cite{cmast:Kohn:2009:JCP} show that {\em extended} SP (XSP) F12 ansatz correlating not only pairs of occupied orbitals but also the occupied-virtual pairs similarly decreases the magnitude of the ground state correlation energy. This defect was addressed by adding an additional singles cluster operator in the more elaborate XSP$_\text{opt}$ approach.\cite{cmast:Kohn:2009:JCP} Curiously, H\"{o}fener et al. decided not to include this additional singles cluster in their formulation of the CC2(F12*)-XSP method,\cite{cmast:Kohn:2019:JCP} perhaps because of the additional rise in complexity and cost due to the need to optimize the extra cluster operator; unfortunately no SP vs. XSP comparison was reported.

 Our experimentation hinted that the F12 geminals with globally fixed geminal lengthscale are suitable for describing correlations of orbitals of the corresponding energy scales. This phenomenon is already well known from the F12 studies of core correlation; the optimal geminal lengthscales for correlation of core orbitals are shorter. Thus, it makes sense that the optimal lengthscale for correlation of high-energy virtual orbitals will be substantially different from that of valence orbitals.
 
 Therefore, we postulated that the optimal use of F12 correlators is to tune them for valence orbitals only. To test this hypothesis, we implemented variants of IP and PP correlators that use a valence subset of virtual orbitals only in their geminal-generating spaces. To distinguish the valence orbitals correlated by the F12 terms from the rest, we refer to them as {\em active} virtuals (not to be confused with the use of term ``active'' in the multiconfiguration context).

The most natural definition of valence virtual orbitals (VVOs) is by projection onto the exact orbitals of ground-state atoms, which for simplicity is emulated by a finite minimal basis set (MBS) of AOs.  Such definition was originally used by Lu et al.\cite{VRG:lu:2004:JCP} and later reformulated in terms of SVD by Schmidt et al.\cite{cmast:Windus:2015:JPCA}
Note that such definition of VVOs is remotely related to the much earlier work of Iwata\cite{VRG:iwata:1981:CPL} and is equivalent (modulo the direct use of SVD vs. an equivalent use of eigendecomposition) to the much more recent AVAS formalism of Sayfutyarova et al.\cite{VRG:sayfutyarova:2017:JCTC} We describe the construction of VVOs here for completeness. Nonorthogonal MBS AOs $\{\ket{\epsilon}\}$ are symmetrically orthonormalized a la L"owdin to produce orthonormal MBS AOs $\{\ket{\overline{\epsilon}}\}$:
  \begin{align}
      \ket{\overline{\epsilon_1}} = \left(\mathbf{s}^{-1/2} \right)^{\epsilon_1}_{\epsilon_2} \ket{\epsilon_2},
  \end{align}
  where $\mathbf{s}$ is the MBS AO metric (overlap matrix). The overlap matrix of the orthonormal MBS AOs with the unoccupied orbitals $\{\ket{e}\}$ that are expanded in OBS AOs $\{\ket{\mu}\}$,
  \begin{align}
      \ket{e} = & C^e_\mu \ket{\mu},
  \end{align}
  is obtained straightforwardfly from
  the overlap of OBS and MBS AOs:
  \begin{align}
  \braket{\overline{\epsilon}}{e} = & \left(\mathbf{s}^{-1/2} \right)^{\epsilon_1}_{\epsilon} \braket{\epsilon_1}{\mu} C^e_\mu 
  \end{align}
 The SVD decomposition of this overlap matrix produces singular values between 1 and 0, with the largest values being very close to 1. The corresponding right singular vectors $\mathbf{V}^\dagger$ (with the rows corresponding to the ket indices $e$) can be interpreted as the linear combinations of the OBS unoccupied orbitals that have the largest overlap with MBS;
 these define the valence/active unoccupied orbitals $\{\ket{a}\}$:
 \begin{align}
     \ket{a} = \left( \mathbf{V}^\dagger \right)^a_e \ket{e}.
 \end{align}
 At physically-relevant geometries, the number of such orbitals $n_a$ equals the rank of MBS minus the number of occupied orbitals. The remaining columns of ${\bf V}^\dagger$ define the inactive unoccupied orbitals $\{\ket{f}\}$.

 Both shapes and energies of VVOs are known to have a well-defined basis set limit, with weak dependence on the basis set as illustrated in \cref{fig:VVO_figure}.

 Whether or not VVOs are used only changes the definition of the active
 orbital space $p'$ (see \cref{figure:spaces}),
 which is used in the definition of PP and IP correlators (\cref{eq:APP,eq:AIP}, respectively). If VVOs are used, space $p'$ is the union of active occupied ($i$) and active/valence unoccupied ($a$) spaces. If VVOs are not used, then $p'$ is the union of active occupied space $i$ and the full OBS unoccupied space $e$.
 Whether or not the geminals are constrained to the valence subspace of virtuals, the expressions for the IP and PP correlators remain unchanged. The projector \cref{eq:Q12} is also unchanged as it includes projection on to the full set of OBS unoccupied orbitals, which preserves the standard expressions for F12 intermediates.
 
\section{Technical Details}\label{sec:methods}
The existing implementation of the transcorrelated F12 Hamiltonian with the II correlator\cite{cmast:Masteran:2023:JCP} in the developmental version of the MPQC package\cite{VRG:peng:2020:JCP} was extended to support the use of IP and PP correlators, with optional restriction of the geminals to the VVO subset of virtuals. In this work, we will refer to results obtained with a transcorrelated Hamiltonian with prefix ``F12-'', appending correlator label as appropriate (e.g., ``F12(OO)-'').

The transcorrelated Hamiltonians are derived using the symbolic tensor algebra package \code{SeQuant}. The resulting tensor algebraic expressions are optimized (``factorized'') and evaluated numerically. The F12 Hamiltonians can then be used as the starting point for a correlated calculation or saved to disk for offline use. The use of IP and PP correlators leads to the appearance of standard intermediates V, X, and B of the F12 methods with geminal ansatz 2 (\cref{eq:V,eq:X,eq:B}) whose robust evaluation using approximate resolution of the identity is well-established\cite{VRG:valeev:2004:CPL,VRG:kedzuch:2005:IJQC}. Whereas the original implementation of the TC F12 Hamiltonian with the II ansatz\cite{cmast:Masteran:2023:JCP} used the existing manually-produced implementation of these intermediates, here the programmable expressions for these intermediates for the II, IP, and PP cases were generated by \code{SeQuant}, which was crucial for extensive experimentation with alternative ansatze that led to this work. The relevant details are discussed further in the Supporting Information.

Ten-no's Slater-type geminal ($-\frac{e^{-\gamma r_{12}}}{\gamma}$) was used as the correlation factor.\cite{VRG:ten-no:2004:CPL}
AO integrals involving the geminal were computed directly, without the usual approximation by a linear combination of Gaussian Geminals.\cite{VRG:may:2004:JCP,VRG:tew:2005:JCP,VRG:may:2005:PCCP}

The standard aug-cc-pV$X$Z (shortened to a$X$Z)\cite{cmast:Dunning:1989},\cite{cmast:Kendall:1992}, their doubly-augmented daug-cc-pV$X$Z variants (shortened to da$X$Z)\cite{cmast:Woon:1994}, and the cc-pV$X$Z-F12 basis sets (shortened to $X$Z-F12) \cite{cmast:Peterson:2008} were used as orbital basis sets (OBS), with the recommended values of geminal exponents  used ($\gamma = \{1.1, 1.2\} a^{-1}_0$ for a\{D,T\}Z OBS, $\gamma = \{0.9, 1.0\} a^{-1}_0$ for \{D,T\}Z-F12 OBS, and $\gamma = \{1.1, 1.3, 1.5\}a^{-1}_0$ for da{D,T,Q}Z OBS). The matching OptRI basis sets of Peterson and co-workers\cite{cmast:Peterson:2009} used to define the CABS space (using the CABS+ approach\cite{VRG:valeev:2004:JCP}) and to approximate the F12 intermediates. All 2-electron integrals were approximated by density fitting in aug-cc-PV$(X+1)$Z-RIFIT \cite{cmast:Weigend:2002} for the corresponding OBS with cardinal number $X$. All VVOs are constructed from the MINI MBS\cite{VRG:andzelm:1984:PSDV1GBSfMC,VRG:vanduijneveldt:1971:}.

The II, IP, and PP correlators used to generate TC Hamiltonians did not correlate core electrons (frozen core approximation). The ground states of the TC F12 Hamiltonian were obtained by complete reoptimization of the Hartree-Fock orbitals with the F12 Hamiltonian (which produces the energy already including some dynamical correlation effects) and the subsequent coupled-cluster singles and doubles method.
The excited states of the TC F12 Hamiltonian were obtained using the frozen-core EOM-CCSD (aka SAC-CI) method\cite{VRG:nakatsuji:1979:,cmast:Stanton1993} whose implementation was previously described.\cite{VRG:peng:2018:JCTC}

Equilibrium bond distances and harmonic vibrational frequencies of excited states were obtained by fitting the potential energy surfaces (PES) evaluated on an equidistant grid with spacing $10^{-3} a_0$ to a 6th-order polynomial, using the equilibrium geometries from Ref. \cite{cmast:Yang:2009} as the guide. Harmonic vibrational frequencies were reported for the most common isotopologue of each species.

\section{Results}\label{sec:results}

\subsection{Vertical Excitation Energies}
\label{sec:results-vee}

 \cref{tab:BH,tab:COH2,tab:N2,tab:CH2,tab:H2O} report the correlation contributions to the electronic energies of ground and excited states as well as vertical excitation energies obtained with standard and F12 TC Hamiltonians for several low-lying singlet states of 5 small molecules: BH, CH$_2$O, N$_2$, H$_2$O, CH$_2$. The BH geometry was taken from Kohn's XSP paper \cite{cmast:Kohn:2009:JCP}, the CH$_2$O geometry was taken from the benchmark study by Schreiber et al.\cite{cmast:Schreiber:2008}, and the remaining three geometries were taken from the HEAT dataset\cite{cmast:Tajti2004}.

Since the F12 geminals only account for the ``2-particle'' basis set incompleteness, conventional F12 approaches require perturbative account for the basis set incompleteness of the ground-state reference using single replacements into the CABS basis\cite{VRG:adler:2007:JCP,VRG:knizia:2008:JCP} (such correction was already introduced in the context of dual basis MP2 method\cite{VRG:wolinski:2003:JCP}). Whereas such a correction has been explored in the context of F12 TC framework before\cite{cmast:Kumar:2022}, our attempts to introduce such a correction in a state-universal manner were not successful, hence here we only focus on the improvements of the dynamical correlation energies that stem from the F12 transcorrelation. An alternative solution to the 1-particle basis set incompleteness problem will be reported elsewhere.

To gain deeper insight into the quality of vertical excitation energies obtained with the F12 transcorrelation Hamiltonians, it is useful to define quantitatively the correlation energies of not only the ground but also the excited states.
For the ground states, we use the the conventional (L\"owdin) correlation energy:
\begin{align}
\label{eq:Ec0}
    E_c^0 
    \equiv & \bra{\bar{\Psi}^0} \hat{\bar{H}} \ket{\bar{\Psi}^0} - \bra{\Phi^0} \hat{H} \ket{\Phi^0},
\end{align}
where $\ket{\Phi^0}$ is the reference wave function. In this work $\bra{\bar{\Psi}^0}$ and $\ket{\bar{\Psi}^0}$ are the left- and right-hand CCSD ground-state wave functions obtained with the F12 TC Hamiltonian.
For well-behaved ground states with the single-configuration character \cref{eq:Ec0} includes primarily the dynamical correlation effects.

No unique excited-state counterpart to \cref{eq:Ec0} is available. Here we defined
the correlation energy of $s$th excited state as
\begin{align}
\label{eq:Ecs}
    E_c^s 
    \equiv & \bra{\bar{\Psi}^s} \hat{\bar{H}} \ket{\bar{\Psi}^s} - \bra{\Phi^s} \hat{H} \ket{\Phi^s},
\end{align}
where $\ket{\Phi^s}$ is the configuration interaction singles (CIS) excited-state wave function and $\bra{\bar{\Psi}^s}$ and $\ket{\bar{\Psi}^s}$ are the left- and right-hand EOM-CCSD wave functions of the $s$th excited state.
\cref{eq:Ecs} is expected to correspond to the dynamical component of the correlation energy of the excited state only if the CIS and EOM-CCSD states are qualitatively the same. Due to this limitation we only reported the excited state correlation energies for the lowest excited state of BH for which CIS and EOM-CCSD descriptions are qualitatively similar and this quantity provided useful insights into the successes and failures of the F12 TC approach.

Our analysis starts with the ground and lowest singlet excited state of BH (see \cref{tab:BH}), which were examined in detail with conventional excited-state CC-F12 approaches by K\"ohn\cite{cmast:Kohn:2009:JCP} and by Ten-no and co-workers\cite{cmast:Ten-no:2010:JCP}.
First, consider the data obtained with the full set of virtual orbitals (FVO). The II correlator (which most resembles the F12 cluster operators utilized in the CC-F12 methods aimed at ground states\cite{VRG:valeev:2008:PCCPP,VRG:adler:2007:JCP,VRG:shiozaki:2008:JCP,VRG:kohn:2008:JCP,VRG:hattig:2010:JCP}) lowers the correlation energy by the equivalent of 2 cardinal quantum numbers. The II correlator also lowers the correlation energy of the excited state, but by a much smaller amount (the aDZ F12 energy is smaller in magnitude than the conventional aTZ counterpart). This is the primary reason why the net effect of the II correlator is to increase the vertical excitation energy, by almost 0.1 eV, rather than decrease it as expected from the basis set convergence of the standard excitation energy. The failure of the II correlator is due to the inability to correlate the single electron that was promoted from the occupied to the virtual orbital with the rest of the electrons.

We expected that the use of IP and PP correlators would resolve the undercorrelation of the excited state. Indeed, they do partially improve the quality of the excitation energy. However, replacing the II correlator by the more flexible IP or PP correlators unexpectedly {\em decreases} the magnitude of the ground state correlation energy, by about 1.5 and 0.5 $mE_\text{h}$ with double and triple-zeta bases, respectively. As we discussed in \cref{sec:formalism}, K\"ohn found similar issues in the context of exploring single-reference F12 coupled-cluster methods for excited states using his XSP, however, this fact is not widely known and we were puzzled by such findings. In fact, the II correlator recovers more excited-state correlation energy at the aTZ level than both the IP and PP correlators using FVOs. When VVOs are used instead, the IP correlator recovers more ground-state correlation energy. This approach also produces improved excitation energies compared to its FVO counterpart. Similar improvement from the use of VVOs is observed for the PP correlator. The VVO PP correlator seems to strike the best balance between ground and excited-state correlation: it recovers almost as much ground state correlation as the II correlator while recovering substantially more correlation energy than the II correlator for the excited state. 

\begin{table}[ht]
\begin{tabular}{l l *{3}{S} *{3}{S}}  
\hline\hline
{} & {} & \multicolumn{3}{c}{FVO} & \multicolumn{3}{c}{VVO} \\ 
\cmidrule(lr){3-5}\cmidrule(lr){6-8}
& & & & & & & \\ 
\multicolumn{1}{c}{Hamiltonian} & \multicolumn{1}{c}{Basis} & 
\multicolumn{1}{c}{$E^0_c$} & \multicolumn{1}{c}{$E^1_c$} & \multicolumn{1}{c}{$\Delta E$} & 
\multicolumn{1}{c}{$E^0_c$} & \multicolumn{1}{c}{$E^1_c$} & \multicolumn{1}{c}{$\Delta E$} \\
\hline
{} & {} & {} & {} & {} & {} & {} & {} \\ 
{} & aDZ & -89.862 & -85.465 & 2.972 & {} & {} & {} \\
$\hat{H}$ & aTZ & -99.214 & -96.741 & 2.930 & {} & {} & {} \\
{} & aQZ & -101.766 & -99.808 & 2.920 & {} & {} & {} \\

{} & {} & {} & {} & {} & {} & {} & {} \\ 

{} & aDZ & -101.219 & -93.492 & 3.062 & {} & {} & {} \\
$\hat{\bar{H}}_{\text{II}}$ & aTZ & -102.757 & -99.881 & 2.941 & {} & {} & {} \\

{} & {} & {} & {} & {} & {} & {} & {} \\ 

{} & aDZ & -99.812 & -95.072 & 2.981 & -100.416 & -96.301 & 2.964 \\
$\hat{\bar{H}}_{\text{IP}}$ & aTZ & -102.161 & -99.491 & 2.935 & -102.444 & -100.404 & 2.918 \\

{} & {} & {} & {} & {} & {} & {} & {} \\ 

{} & aDZ & -99.867 & -95.194 & 2.979 & -100.902 & -96.825 & 2.963 \\
$\hat{\bar{H}}_{\text{PP}}$ & aTZ & -102.330 & -99.534 & 2.939 & -102.719 & -100.465 & 2.924 \\

{} & {} & {} & {} & {} & {} & {} & {} \\ 
\hline\hline
\end{tabular}
\caption{Comparison of ground state and lowest lying excited singlet state
correlation energies between the various ansatze for BH system. Correlation energies are reported in millihartree (m$E_h$), vertical excitation energies are reported in electron volts (eV).}.
\label{tab:BH}
\end{table}

\begin{table}[ht]
    \begin{tabular}{lccccccccccc}
    \hline\hline
          {} & {} & \multicolumn{5}{c}{FVO} & \multicolumn{5}{c}{VVO} \\ \cmidrule(lr){3-7}\cmidrule(lr){8-12}
 &  &  &  &  &  &  &  \\ 
\multicolumn{1}{c}{Hamiltonian} & \multicolumn{1}{c}{Basis} &  \multicolumn{1}{c}{$E^0_{c}$ } & \multicolumn{1}{c}{S$_1$} & \multicolumn{1}{c}{S$_2$} & \multicolumn{1}{c}{S$_3$ } &  \multicolumn{1}{c}{S$_4$ } &\multicolumn{1}{c}{$E^0_c$ } & \multicolumn{1}{c}{S$_1$} & \multicolumn{1}{c}{S$_2$} & \multicolumn{1}{c}{S$_3$ } & \multicolumn{1}{c}{S$_4$ }\\
\hline
{} & {} & {} & {} & {} & {} & {} & {} & \\ 
{} & aDZ &-349.125  & 3.943 &7.053 & 7.989 &8.068 &\\
{$\hat{H}$} & aTZ & -413.104 & 3.935 & 7.239 & 8.114& 8.225& \\
{} & aQZ & -433.832 & 3.946 & 7.304 & 8.148 & 8.281 &
{} & {} & {} & {} & {} \\
{} & {} & {} & {} & {} & {} & {} & {} & \\ 
{} & aDZ & -437.714 & 4.312 & 7.292 & 8.237 & 8.307 \\
$\hat{\bar{H}}_{\text{II}}$ & aTZ & -443.980 & 4.048 & 7.305 & 8.184 & 8.291\\
 {} & {} & {} & {} & {} & {} & {} & {} & \\ 
 {} & aDZ & -438.125 & 4.010&7.272&8.208 & 8.295 &-438.058&3.985&7.285&8.251& 8.303\\
 $\hat{\bar{H}}_{\text{IP}}$ & aTZ & -443.384&4.001&7.311&8.189&8.296& -443.998& 3.948 & 7.311 & 8.196 & 8.295 \\
 {} & {} & {} & {} & {} & {} & {} & {} &  {} & \\
{} & aDZ & -436.516 & 4.013 & 7.266 & 8.204 & 8.288 &-438.487& 4.006 & 7.287 & 8.256 & 8.307 \\
 $\hat{\bar{H}}_{\text{PP}}$ & aTZ &-442.941 &3.971 & 7.312&8.190 &8.299 & -444.366& 3.960 & 7.313 & 8.200 & 8.298 \\\\ 
 \hline \hline
    \end{tabular}
    \caption{Comparing ground state correlation energy and vertical excitation energies generated by various ansatze for CH$_2$O system. Correlation energy in  millihartree (m$E_h$), gaps are reported in eV.}
    \label{tab:COH2}
\end{table}

All correlators significantly lower the ground state correlation energies of the formaldehyde molecule.\cref{tab:COH2}. The FVO vs. VVO contrast is especially striking here: in most instances the FVO-based correlators produce smaller correlation energy than the VVO counterpart. And again, the PP VVO correlator produces the lowest correlation energies with both aDZ and aTZ bases.

Like the case of BH where the lowest excited state was valence in character and therefore the excitation energy exhibited little basis set variation, the lowest excited state of formaldehyde has a valence character also, with nearly nonexistent dependence on the basis. As expected, the II correlator significantly overestimates this state's excitation energy by 0.35 eV at the aDZ level by undercorrelating the excited electron. The use of IP and PP correlators addresses this issue nicely, with the VVO variants again superior to the FVO counterparts.
Unlike the lowest excited state, the higher-lying S states of formaldehyde have a substantial Rydberg character thereby exhibiting a more protracted basis set convergence, with double-zeta excitation energies in error by 0.2-0.3 eV and converging from below. The use of VVO IP and PP correlators significantly reduces the basis set errors of the excitation energies for these 3 states, especially with the double-zeta basis. The $\mathrm{S}_3$ state is the only instance where the DZ and TZ F12 excitation energies differ by more than a couple of dozens of mEv; this may suggest substantial 1-particle basis set incompleteness effects on the excitation energy of this state that the F12 transcorrelation does not incorporate in its present form.

\begin{table}[ht]
    \begin{tabular}{lcccccccccc}
    \hline\hline
     {} & {} & \multicolumn{4}{c}{FVO} & \multicolumn{4}{c}{VVO} \\ \cmidrule(lr){3-6}\cmidrule(lr){7-10}
 &  &  &  &  &  &  &  \\ 
\multicolumn{1}{c}{Hamiltonian} & \multicolumn{1}{c}{Basis} &  \multicolumn{1}{c}{$E^0_c$ } & \multicolumn{1}{c}{S$_1$} & \multicolumn{1}{c}{S$_2$} & \multicolumn{1}{c}{S$_3$ } &\multicolumn{1}{c}{$E^0_c$ } & \multicolumn{1}{c}{S$_1$} & \multicolumn{1}{c}{S$_2$} & \multicolumn{1}{c}{S$_3$ } & \\
\hline
{} & {} & {} & {} & {} & {} & {} & {} & \\ 
{} & aDZ & -319.811 & 9.537 & 10.258 & 10.666  & \\
 $\hat{H}$ & aTZ & -376.877 &9.451 &10.057&10.503 \\
{} & aQZ & -395.236 & 9.442 & 10.042 & 10.482 & \\
{} & {} & {} & {} & \\ 

{} & aDZ &-399.309 &9.925&10.699&10.988  \\
$\hat{\bar{H}}_{\text{II}}$ & aTZ &-404.694&9.567&10.203&10.592& \\
{} & {} &  &  & \\ 
{} & aDZ & -399.646 & 9.589& 10.297& 10.698 &-399.690&9.553&10.265&10.659& \\
{$\hat{\bar{H}}_{\text{IP}}$} & aTZ & -404.265 &9.480&10.089&10.524&-404.661&9.451&10.063&10.492 \\
{} & {} & {} & {} & \\ 
{} & aDZ &-398.268&9.577&10.313&10.702&-400.112&9.560&10.307&10.681& \\
 $\hat{\bar{H}}_{\text{PP}}$ & aTZ & -403.785 &9.483&10.098&10.528&-405.066&9.464&10.090&10.508& \\
{} & {} & {} & {} & \\ 
\hline\hline
    \end{tabular}
    \caption{N$_2$ singlet ground state and first few excited singlet states. Results using FVO and VVO}
    \label{tab:N2}
\end{table}

Our findings are similar for the N$_2$ molecule (\cref{tab:N2}) in that the VVO IP and PP correlators provide the best overall accuracy for the ground and excited states. The II correlator overestimates the excitation energies by as much as 0.5 eV, with the VVO IP and PP correlators reducing the error. However, unlike the case of formaldehyde, where the excitation energies of Rydberg-character states converged from below, here the excitation energies of all states converge from above. This again suggests the interplay between the 1- and 2-particle basis set incompleteness effects; this hypothesis is supported by the fact that the F12 transcorrelation (with the VVO IP and PP correlators) seems to have a relatively minor effect on the excitation energies; it is likely that the residual 1-particle basis set effects are important.

\begin{table}[ht]
    \begin{tabular}{lcccccccccc}
          \hline\hline
     {} & {} & \multicolumn{4}{c}{FVO} & \multicolumn{4}{c}{VVO} \\ \cmidrule(lr){3-6}\cmidrule(lr){7-10}
 &  &  &  &  &  &  &  \\ 
\multicolumn{1}{c}{Hamiltonian} & \multicolumn{1}{c}{Basis} &  \multicolumn{1}{c}{S$_0$ } & \multicolumn{1}{c}{S$_1$} & \multicolumn{1}{c}{S$_2$} & \multicolumn{1}{c}{S$_3$ } &\multicolumn{1}{c}{S$_0$ } & \multicolumn{1}{c}{S$_1$} & \multicolumn{1}{c}{S$_2$} & \multicolumn{1}{c}{S$_3$ } & \\
\hline
{}&{}&{}&{}&{}&{}&{}&{}&{}&{}\\
{}&DZ-F12&-154.014&1.718&5.757&6.139&\\
{$\hat{H}$}&TZ-F12 & -167.784&1.679&5.716&6.257&\\
{}&QZ-F12&-172.195&1.673&5.716&6.292&\\
{}&{}&{}&{}&{}&{}&{}&{}&{}&{}\\
{}&DZ-F12&
-172.867
 &
1.859
&
5.923
&
6.334
&\\
{$\hat{\bar{H}}_{\text{II}}$}&TZ-F12&-175.401&1.718&5.772&6.312&&\\
{}&{}&{}&{}&{}&{}&{}&{}&{}&{}\\

{}&DZ-F12&-172.061&1.753&5.802&6.169&-172.487&1.735&5.785&6.182\\
{$\hat{\bar{H}}_{\text{IP}}$}&TZ-F12&-174.932&1.696&5.743&6.253&-175.209&1.681&5.730&6.258\\
{}&{}&{}&{}&{}&{}&{}&{}&{}&{}\\
{}&DZ-F12&-171.768&1.765&5.814&6.090&-172.932&1.755&5.808&6.118&\\
{$\hat{\bar{H}}_{\text{PP}}$}&TZ-F12 &-174.873&1.702&5.747&6.225&-175.451&1.692&5.740&6.236&\\
\hline \hline
    \end{tabular}
    \caption{CH$_2$ singlet ground state and first few excited singlet states. results using FVO and VVO.}
    \label{tab:CH2}
\end{table}

Only minor differences with the previous cases were observed for methylene (\cref{tab:CH2}). The use of cc-pVXZ-F12 basis sets purpose-built for F12 methods\cite{cmast:Peterson:2008} does not change the rate of convergence or magnitude of F12-spurred improvements: the ground-state energy again is improved by the equivalent of more than 2 cardinal basis set numbers. Out of the 3 excited states, excitation energies of two (S$_1$ and S$_2$) converge from above (like in N$_2$) and one converges from below (like in CH$_2$O). Thus, as expected, F12 transcorrelation (with VVO IP and PP correlators) does not seem to produce substantial effects on the excitation energies of the former and, somewhat unexpectedly, does not show substantial improvements for the latter. The interplay of the 1- and 2-particle basis set incompleteness effects is likely the cause of these differences.

\begin{table}[ht]
\begin{tabular}{l l l l l l l l l l}   
    \hline\hline
 &  & \multicolumn{4}{l}{FVO} & \multicolumn{4}{l}{VVO} \\ \cmidrule(lr){3-6}\cmidrule(lr){7-10}
 &  &  &  &  &  &  &  &  &  \\
Hamiltonian & Basis & S$_0$ & S$_1$ & S$_2$ & S$_3$ & S$_0$ & S$_1$ & S$_2$ & S$_3$ \\
\hline
 &  &  &  &  &  &  &  &  &  \\
 & aDZ & -230.040 & 7.082 & 8.847 & 8.923 &  &  &  &  \\
$\hat{H}$ & aTZ & -275.393 & 7.215 & 8.985 & 9.018 &  &  &  &  \\
 & aQZ & -290.443 & 7.277 & 9.045 & 9.070 &  &  &  &  \\
 &  &  &  &  &  &  &  &  &  \\
$\hat{\bar{H}}_{\text{II}}$ & aDZ & -293.187 & 7.341 & 9.104 & 9.148 &  &  &  &  \\
 & aTZ & -297.276 & 7.293 & 9.062 & 9.083 &  &  &  &  \\
 &  &  &  &  &  &  &  &  &  \\
$\hat{\bar{H}}_{\text{IP}}$ & aDZ & -293.310 & 7.287 & 9.064 & 9.101 & -293.086 & 7.299 & 9.072 & 9.116 \\
 & aTZ & -296.955 & 7.293 & 9.064 & 9.084 & -297.185 & 7.292 & 9.060 & 9.085 \\
 &  &  &  &  &  &  &  &  &  \\
$\hat{\bar{H}}_{\text{PP}}$ & aDZ & -292.421 & 7.285 & 9.059 & 9.098 & -293.328 & 7.307 & 9.081 & 9.119 \\
 & aTZ & -296.489 & 7.293 & 9.063 & 9.083 & -297.331 & 7.298 & 9.065 & 9.088 \\
\hline \hline
\end{tabular}
\caption{H$_2$O singlet ground state and first few excited singlet states. Results using FVO and VVO}
    \label{tab:H2O}
\end{table}

Finally, the water molecule (\cref{tab:H2O}) exhibits patterns similar to those of the other systems. With the excitation energies converging from below, the F12 transcorrelation greatly helps with the basis set convergence. The VVO PP correlator again provides the most balanced performance.

\subsection{Excited State Properties}
\label{sec:results-esprop}

\begin{table}[ht!]
\begin{tabular}{l l| l l l l l| l l | l l | l l} 
\hline\hline
 & & \multicolumn{5}{l}{standard$^a$} & \multicolumn{2}{l}{F12(II)} & \multicolumn{2}{l}{ F12(VVO PP)} & \multicolumn{2}{l}{(F12)$^b$} \\
 \hline
 & basis & daDZ & daTZ & daQZ & da5Z & CBS & daTZ & daQZ &  daTZ & daQZ & daTZ & daQZ\\
 \hline
 &  &  &  &  &  &  &  &  &  &  &  &    \\
\multirow{5}{*}{N$_2$} & $a'$ $^1\Sigma_u^-$ & 126.7 & 125.4 & 125.1 & 125 & 124.9 & 125.1 & 124.9 &  125.1 & 124.9 & 125.1 & 124.9 \\
 &  &  &  &  &  &  &  &  &  &  &  &    \\
 & $a$ $^1\Pi_g$ & 122 & 120.7 & 120.3 & 120.2 & 120.1 &  120.4 & 120.2 &  120.4 & 120.1 & 120.4 & 120.2 \\
 &  &  &  &  &  &  &  &  &  &  &  &    \\
 & $w$ $^1\Delta_u$ & 126.2 & 124.9 & 124.5 & 124.4 & 124.4 &  124.6 & 124.4 & 124.6 & 124.4 & 124.6 & 124.4 \\
 \hline
 &  &  &  &  &  &  &  &  &  &     \\
\multirow{5}{*}{CO} & $A$ $^1\Pi$ & 124.5 & 123.1 & 122.5 & 122.4 & 122.3 &  122.7 & 122.3 &  122.7 & 122.4 & 122.6 & 122.3 \\
 &  &  &  &  &  &  &  &  &  &     \\
 & $B$ $^1\Sigma^+$ & 112.7 & 111.6 & 111.2 & 111.1 & 111 &  111.3 & 111 & 111.3 & 111 & 111.2 & 111 \\
 &  &  &  &  &  &  &  &  &  &     \\
 & $C$ $^1\Sigma^+$ & 112.6 & 111.5 & 111 & 110.9 & 110.8 &  111.1 & 110.9 &  111.1 & 110.9 & 111.1 & 110.9 \\
 \hline
 &  &  &  &  &  &  &  &  &  &      \\
\multirow{5}{*}{BF} & $A$ $^1\Pi$ & 135.1 & 131.1 & 130.5 & 130.3 & 130.2 &  130.7 & 130.3 & 130.7 & 130.3 & 130.6 & 130.3 \\
 &  &  &  &  &  &  &  &  &  &      \\
 & $B$ $^1\Sigma^+$ & 123.7 & 121.2 & 120.7 & 120.6 & 120.5 &  120.9 & 120.6 & 121 & 120.6 & 120.9 & 120.6 \\
 &  &  &  &  &  &  &  &  &  &      \\
 & $C$ $^1\Sigma^+$ & 125.2 & 122.5 & 122 & 121.9 & 121.8 &  122.2 & 121.9 & 122.2 & 121.9 & 122.2 & 121.9 \\
 \hline
 &  &  &  &  &  &  &  &  &  &      \\
\multirow{3}{*}{BH} & $A$ $^1\Pi$ & 124.4 & 122.4 & 122.2 & 122.1 & 122.1 &  122.2 & 122.1 & 122.2 & 122.1 & 122.3 & 122.1 \\
 &  &  &  &  &  &  &  &  &  &      \\
 & $B$ $^1\Sigma^+$ & 123.4 & 121.8 & 121.6 & 121.6 & 121.5 & 121.7 & 121.6 & 121.7 & 121.6 & 121.7 & 121.6 \\
 \hline\hline
\label{tab:eqr}
\end{tabular}

$^a$ Standard EOM-CCSD results from Ref. \citenum{cmast:Yang:2009}.\\
$^b$ EOM-CCSD(F12) results from Ref. \citenum{cmast:Yang:2009}.

\caption{Excited state equilibrium bond lengths (pm) for lowest singlet excited states of several small diatomics. F12-transcorrelated EOM-CCSD approaches with II and VVO PP correlators are contrasted with the standard and conventional F12-correlated EOM-CCSD formalisms.}
\end{table}

Unlike the excitation energies, which are often relatively weakly dependent on the basis set due to the frequently excellent cancellation of the basis set incompleteness errors between low-lying states, properties of individual excited states are just as sensitive to the basis set effects as their ground state counterparts. To probe the ability of the F12 TC approaches to reduce the basis set errors of excited state properties, we examined equilibrium geometries and harmonic vibrational frequencies of several paradigmatic  diatomics (N$_2$, CO, BF, and BH) obtained with standard explicitly-correlated variants of EOM-CCSD energies. The F12 transcorrelated EOM-CCSD results obtained with the II  and VVO-only form of PP correlators were compared against the standard EOM-CCSD and the conventional F12 form of the EOM-CCSD due to Yang and H\"attig in which F12 terms are incorporated into the ground state clusters and the excited state wave vectors.\cite{cmast:Yang:2009} Note that the EOM-CCSD(F12) approach of Yang and H\"attig uses the older orbital-invariant ansatz of the F12 theory\cite{VRG:klopper:1991:CPL} in which each pair of orbitals is correlated by an adjustable superposition of geminals, rather than a single fixed-parameter geminal\cite{VRG:ten-no:2004:JCP} typically used in modern ground-state F12 methods as well as in the F12 transcorrelated approaches described here. The results are reported in \cref{tab:eqr} and \cref{tab:freq}, respectively.

The basis set convergence of equilibrium bond distances (see \cref{tab:eqr}) is substantially improved by the addition of explicitly-correlated terms; on average the F12 approaches gain the equivalent of a single cardinal number in the orbital basis. The equilibrium bond distances obtained with II and VVO PP variants of the F12 transcorrelation are almost indistinguishable, in contrast to their behavior for correlation energies and the excitation energies discussed earlier. Both transcorrelated approaches produce bond distances that are nearly identical to those from the traditional F12 variant of the EOM-CCSD.

The basis set convergence of harmonic vibrational frequencies is also significantly improved by the F12 explicit correlation; again, on average the F12 approaches gain the equivalent of a single cardinal number in the orbital basis. Unlike the minute differences between the use of II and VVO PP correlators for the equilibrium geometries, there are more noticeable differences between the two correlators, with the PP correlator producing smaller basis set errors. For example, with the daQZ basis the \{mean,max\} absolute basis set errors obtained with the II and VVO PP correlators are \{2.1,6\} cm$^{-1}$ and \{2.1,5\} cm$^{-1}$ respectively. For the majority of states the frequencies obtained with the PP transcorrelator match up well with those from the conventional (F12) variant of EOM-CCSD, but
overall the latter is more precise, with the \{mean,max\} absolute basis set errors of \{0.7,3\} cm$^{-1}$ with the daQZ basis. This result is not unexpected as the H{\"a}ttig and Yang implementation is more robust likely due to the more flexible F12 ansatz including many adjustable parameters rather than the fixed amplitude SP ansatz used in the transcorrelated F12 approaches.
Note that we recently discovered that the recommended F12 geminal exponents are better suited for the older orbital-invariant CC F12 methods than their SP-ansatz counterparts\cite{cmast:Powell:2025}. The use of updated geminal exponents with the SP-ansatz-based transcorrelation may help further improve the reported performance relative to that of EOM-CCSD(F12).

\begin{table}[ht!]
\begin{tabular}{l l| l l l l l| l l| l l| l l} 
\hline \hline
 & & \multicolumn{5}{l}{standard$^a$} & \multicolumn{2}{l}{F12(II)} & \multicolumn{2}{l}{ F12(VVO PP)} & \multicolumn{2}{l}{(F12)$^b$} \\
 \hline
 & basis & daDZ & daTZ & daQZ & da5Z & CBS & daTZ & daQZ &  daTZ & daQZ & daTZ & daQZ\\
 &  &  &  &  &  &  &  &  &  &  &  &    \\
\multirow{5}{*}{N$_2$} & $a'$ $^1\Sigma_u^-$  & 1709 & 1714 & 1725 & 1727 & 1730 & 1722 &1727 &1723&1728 & 1724 & 1729 \\
 &  &  &  &  &  &  &  &  &  &  &  &   \\
 & $a$ $^1\Pi_g$ & 1827 & 1830 & 1848 & 1851 & 1854 & 1846& 1854 & 1846 & 1853 & 1847 & 1854 \\
 &  &  &  &  &  &  &  &  &  &  &  &    \\
 & $w$ $^1\Delta_u$ & 1726 & 1733 & 1744 & 1746 & 1749 &1742 & 1748 & 1744 & 1748 & 1744 & 1748 \\
 \hline
 &  &  &  &  &  &  &  &  &  &  &  &    \\
\multirow{5}{*}{CO} & $A$ $^1\Pi$ & 1510 & 1561 & 1581 & 1587 & 1592 &1574 & 1586 &1574 &1587& 1580 & 1589 \\
 &  &  &  &  &  &  &  &  &  &  &  &    \\
 & $B$ $^1\Sigma^+$ & 2184 & 2221 & 2244 & 2249 & 2253 &2241&2250&2241&2249 & 2246 & 2253 \\
 &  &  &  &  &  &  &  &  &  &  &  &    \\
 & $C$ $^1\Sigma^+$ & 2219 & 2259 & 2281 & 2285 & 2289 &2277&2287&2276&2286 & 2282 & 2289 \\
 \hline
 &  &  &  &  &  &  &  &  &  &  &  &    \\
\multirow{5}{*}{BF} & $A$ $^1\Pi$ & 1127 & 1258 & 1271 & 1275 & 1279 &1267&1274&1272&1276& 1275 & 1278 \\
 &  &  &  &  &  &  &  &  &  &  &  &    \\
 & $B$ $^1\Sigma^+$ & 1580 & 1704 & 1715 & 1718 & 1720 & 1715&1719& 1712&1718 & 1717 & 1720 \\
 &  &  &  &  &  &  &  &  &  &  &  &    \\
 & $C$ $^1\Sigma^+$ & 1487 & 1620 & 1632 & 1635 & 1637 &1632&1636&1639&1635& 1633 & 1637 \\
 \hline
 &  &  &  &  &  &  &  &  &  &  &  &    \\
\multirow{3}{*}{BH} & $A$ $^1\Pi$ & 2229 & 2303 & 2321 & 2323 & 2326 &2327& 2325& 2321 &2326 &2314 & 2324 \\
 &  &  &  &  &  &  &  &  &  &  &  &    \\
 & $B$ $^1\Sigma^+$ & 2360 & 2390 & 2398 & 2399 & 2400 & 2403 & 2400 &  2398 & 2400 & 2398 & 2400 \\
\hline\hline
\end{tabular}

$^a$ Standard EOM-CCSD results from Ref. \citenum{cmast:Yang:2009}.\\
$^b$ EOM-CCSD(F12) results from Ref. \citenum{cmast:Yang:2009}.

\label{tab:freq}
\caption{Harmonic vibrational frequencies (cm$^{-1}$) for the lowest singlet excited states of several small diatomics. F12-transcorrelated EOM-CCSD approaches with II and VVO PP correlators are contrasted with the standard and conventional F12-correlated EOM-CCSD formalisms.}
\end{table}

\section{Summary} \label{sec:summary}
The approximate unitary F12-style transcorrelation\cite{VRG:yanai:2012:JCP,cmast:Masteran:2023:JCP} is a promising approach to address the basis set incompleteness error of correlated electronic states and their properties that produces Hermitian effective Hamiltonians with 2-particle interactions only. The Hermitian and non-Hermitian\cite{VRG:ten-no:2023:JCP} F12 transcorrelation frameworks share the core design trait with the traditional F12 models\cite{VRG:klopper:2006:IRPC,VRG:kong:2012:CR,VRG:hattig:2012:CR,VRG:ten-no:2012:TCA} that only the universal (thereby, necessarily, spin-dependent) short-range correlation physics is encoded by a fixed 1-parameter correlator, with the rest of system and state-specific correlation effects modeled by traditional Fock-space expansions.

In this contribution we explored whether it is possible to use F12 transcorrelation for balanced description of ground and low-energy excited states using single-reference coupled-cluster methods. Traditional (non-transcorrelated) explicitly-correlated F12 methods capable of treating excited states have been designed before by extending the traditional II ansatz that uses F12 geminals for correlation of pairs of occupied reference orbitals only (thereby undercorrelating the excited states\cite{cmast:Fliegl:2006:JCP}) to include correlation of occupied-virtual reference pairs (IP ansatz).\cite{cmast:Neiss:2006:JCP,VRG:neiss:2007:JCP,cmast:Kohn:2009:JCP,cmast:Kohn:2019:JCP,cmast:Ten-no:2010:JCP,cmast:Bartlett:2015:JCP} We found that the straightforward extension of the F12 transcorrelation to correlate virtual orbitals results in an unphysical undercorrelation of the ground state, but constraining the approach to use valence-like virtuals only (VVOs) solves this issue and leads to a balanced and accurate description of the ground and low-energy excited states of small systems.

The F12 transcorrelated Hamiltonian based on the VVO PP correlator (in which all pairs involving occupied and valence virtual orbitals are correlated) robustly reduces the basis set errors in correlation energies and in vertical excitation energies (most pronounced for states with Rydberg character). However, for some excited states the residual 1-particle basis set errors can be substantial and will require further refinements of the approach. The properties of individual excited states, such as geometries and especially vibrational frequencies, are also substantially improved by the F12 transcorrelation.

Improvements of the approach focusing on the residual 1-particle basis set incompleteness effects, extensions to special relativity, and assessment of the performance for response properties of molecules will be reported elsewhere.

\begin{suppinfo}

Cartesian geometries of the systems studied in \cref{sec:results-vee}, the equations for transcorrelated Hamiltonian obtained with the IP correlator (\cref{eq:AIP}) which have not been reported previously.

\end{suppinfo}

\begin{acknowledgement}
This research was supported by the US Department of Energy, Office of Science, via award DE-SC0022327. The development of the \code{SeQuant} software library is supported by the US National Science Foundation via award 2217081. The authors acknowledge Advanced Research Computing at Virginia Tech (\url{https://arc.vt.edu/}) for providing computational resources and technical support that have contributed to the results reported within this paper.
\end{acknowledgement}

\bibliography{vrgrefs,cmastrefs}

\providecommand{\latin}[1]{#1}
\makeatletter
\providecommand{\doi}
  {\begingroup\let\do\@makeother\dospecials
  \catcode`\{=1 \catcode`\}=2 \doi@aux}
\providecommand{\doi@aux}[1]{\endgroup\texttt{#1}}
\makeatother
\providecommand*\mcitethebibliography{\thebibliography}
\csname @ifundefined\endcsname{endmcitethebibliography}
  {\let\endmcitethebibliography\endthebibliography}{}
\begin{mcitethebibliography}{92}
\providecommand*\natexlab[1]{#1}
\providecommand*\mciteSetBstSublistMode[1]{}
\providecommand*\mciteSetBstMaxWidthForm[2]{}
\providecommand*\mciteBstWouldAddEndPuncttrue
  {\def\EndOfBibitem{\unskip.}}
\providecommand*\mciteBstWouldAddEndPunctfalse
  {\let\EndOfBibitem\relax}
\providecommand*\mciteSetBstMidEndSepPunct[3]{}
\providecommand*\mciteSetBstSublistLabelBeginEnd[3]{}
\providecommand*\EndOfBibitem{}
\mciteSetBstSublistMode{f}
\mciteSetBstMaxWidthForm{subitem}{(\alph{mcitesubitemcount})}
\mciteSetBstSublistLabelBeginEnd
  {\mcitemaxwidthsubitemform\space}
  {\relax}
  {\relax}

\bibitem[Hylleraas(1929)]{VRG:hylleraas:1929:ZFP}
Hylleraas,~E.~A. {Neue Berechnung der Energie des Heliums im Grundzustande,
  sowie des tiefsten Terms von Ortho-Helium}. \emph{Z. Physik} \textbf{1929},
  \emph{54}, 347--366\relax
\mciteBstWouldAddEndPuncttrue
\mciteSetBstMidEndSepPunct{\mcitedefaultmidpunct}
{\mcitedefaultendpunct}{\mcitedefaultseppunct}\relax
\EndOfBibitem
\bibitem[Slater(1928)]{VRG:slater:1928:PRa}
Slater,~J.~C. The {{Normal State}} of {{Helium}}. \emph{Phys. Rev.}
  \textbf{1928}, \emph{32}, 349--360\relax
\mciteBstWouldAddEndPuncttrue
\mciteSetBstMidEndSepPunct{\mcitedefaultmidpunct}
{\mcitedefaultendpunct}{\mcitedefaultseppunct}\relax
\EndOfBibitem
\bibitem[Slater(1928)]{VRG:slater:1928:PRb}
Slater,~J.~C. Central {{Fields}} and {{Rydberg Formulas}} in {{Wave
  Mechanics}}. \emph{Phys. Rev.} \textbf{1928}, \emph{31}, 333--343\relax
\mciteBstWouldAddEndPuncttrue
\mciteSetBstMidEndSepPunct{\mcitedefaultmidpunct}
{\mcitedefaultendpunct}{\mcitedefaultseppunct}\relax
\EndOfBibitem
\bibitem[Kato(1957)]{VRG:kato:1957:CPAM}
Kato,~T. On the Eigenfunctions of Many-particle Systems in Quantum Mechanics.
  \emph{Commun. Pure Appl. Math.} \textbf{1957}, \emph{10}, 151--177\relax
\mciteBstWouldAddEndPuncttrue
\mciteSetBstMidEndSepPunct{\mcitedefaultmidpunct}
{\mcitedefaultendpunct}{\mcitedefaultseppunct}\relax
\EndOfBibitem
\bibitem[Pack and Byers~Brown(1966)Pack, and Byers~Brown]{VRG:pack:1966:JCP}
Pack,~R.~T.; Byers~Brown,~W. Cusp Conditions for Molecular Wavefunctions.
  \emph{J. Chem. Phys.} \textbf{1966}, \emph{45}, 556\relax
\mciteBstWouldAddEndPuncttrue
\mciteSetBstMidEndSepPunct{\mcitedefaultmidpunct}
{\mcitedefaultendpunct}{\mcitedefaultseppunct}\relax
\EndOfBibitem
\bibitem[Ruiz \latin{et~al.}(2021)Ruiz, Sims, and Padhy]{VRG:ruiz:2021:AiQC}
Ruiz,~M.~B.; Sims,~J.~S.; Padhy,~B. \emph{Advances in {{Quantum Chemistry}}};
  Elsevier, 2021; Vol.~83; pp 171--208\relax
\mciteBstWouldAddEndPuncttrue
\mciteSetBstMidEndSepPunct{\mcitedefaultmidpunct}
{\mcitedefaultendpunct}{\mcitedefaultseppunct}\relax
\EndOfBibitem
\bibitem[Mitroy \latin{et~al.}(2013)Mitroy, Bubin, Horiuchi, Suzuki, Adamowicz,
  Cencek, Szalewicz, Komasa, Blume, and Varga]{VRG:mitroy:2013:RMP}
Mitroy,~J.; Bubin,~S.; Horiuchi,~W.; Suzuki,~Y.; Adamowicz,~L.; Cencek,~W.;
  Szalewicz,~K.; Komasa,~J.; Blume,~D.; Varga,~K. Theory and Application of
  Explicitly Correlated {{Gaussians}}. \emph{Rev. Mod. Phys.} \textbf{2013},
  \emph{85}, 693--749\relax
\mciteBstWouldAddEndPuncttrue
\mciteSetBstMidEndSepPunct{\mcitedefaultmidpunct}
{\mcitedefaultendpunct}{\mcitedefaultseppunct}\relax
\EndOfBibitem
\bibitem[Nakatsuji(2004)]{VRG:nakatsuji:2004:PRL}
Nakatsuji,~H. Scaled {{Schr{\"o}dinger Equation}} and the {{Exact Wave
  Function}}. \emph{Phys. Rev. Lett.} \textbf{2004}, \emph{93}, 030403\relax
\mciteBstWouldAddEndPuncttrue
\mciteSetBstMidEndSepPunct{\mcitedefaultmidpunct}
{\mcitedefaultendpunct}{\mcitedefaultseppunct}\relax
\EndOfBibitem
\bibitem[Foulkes \latin{et~al.}(2001)Foulkes, Mitas, Needs, and
  Rajagopal]{VRG:foulkes:2001:RMP}
Foulkes,~W.; Mitas,~L.; Needs,~R.; Rajagopal,~G. Quantum Monte Carlo
  Simulations of Solids. \emph{Rev. Mod. Phys.} \textbf{2001}, \emph{73},
  33--83\relax
\mciteBstWouldAddEndPuncttrue
\mciteSetBstMidEndSepPunct{\mcitedefaultmidpunct}
{\mcitedefaultendpunct}{\mcitedefaultseppunct}\relax
\EndOfBibitem
\bibitem[Boys(1969)]{VRG:boys:1969:PRSLA}
Boys,~S.~F. Some Bilinear Convergence Characteristics of the Solutions of
  Dissymmetric Secular Equations. \emph{Proc. R. Soc. Lond. A} \textbf{1969},
  \emph{309}, 195--208\relax
\mciteBstWouldAddEndPuncttrue
\mciteSetBstMidEndSepPunct{\mcitedefaultmidpunct}
{\mcitedefaultendpunct}{\mcitedefaultseppunct}\relax
\EndOfBibitem
\bibitem[Boys and Handy(1969)Boys, and Handy]{VRG:boys:1969:PRSMPES}
Boys,~S.~F.; Handy,~N.~C. A Condition to Remove the Indeterminacy in
  Interelectronic Correlation Functions. \emph{Proc. R. Soc. Math. Phys. Eng.
  Sci.} \textbf{1969}, \emph{309}, 209--220\relax
\mciteBstWouldAddEndPuncttrue
\mciteSetBstMidEndSepPunct{\mcitedefaultmidpunct}
{\mcitedefaultendpunct}{\mcitedefaultseppunct}\relax
\EndOfBibitem
\bibitem[Kutzelnigg(1985)]{VRG:kutzelnigg:1985:TCA}
Kutzelnigg,~W. R 12-{{Dependent}} Terms in the Wave Function as Closed Sums of
  Partial Wave Amplitudes for Large l. \emph{Theor. Chim. Acta} \textbf{1985},
  \emph{68}, 445--469\relax
\mciteBstWouldAddEndPuncttrue
\mciteSetBstMidEndSepPunct{\mcitedefaultmidpunct}
{\mcitedefaultendpunct}{\mcitedefaultseppunct}\relax
\EndOfBibitem
\bibitem[Kong \latin{et~al.}(2012)Kong, Bischoff, and Valeev]{VRG:kong:2012:CR}
Kong,~L.; Bischoff,~F.~A.; Valeev,~E.~F. Explicitly Correlated {{R12}}/{{F12}}
  Methods for Electronic Structure. \emph{Chem. Rev.} \textbf{2012},
  \emph{112}, 75--107\relax
\mciteBstWouldAddEndPuncttrue
\mciteSetBstMidEndSepPunct{\mcitedefaultmidpunct}
{\mcitedefaultendpunct}{\mcitedefaultseppunct}\relax
\EndOfBibitem
\bibitem[H{\"a}ttig \latin{et~al.}(2012)H{\"a}ttig, Klopper, K{\"o}hn, and
  Tew]{VRG:hattig:2012:CR}
H{\"a}ttig,~C.; Klopper,~W.; K{\"o}hn,~A.; Tew,~D.~P. Explicitly Correlated
  Electrons in Molecules. \emph{Chem. Rev.} \textbf{2012}, \emph{112},
  4--74\relax
\mciteBstWouldAddEndPuncttrue
\mciteSetBstMidEndSepPunct{\mcitedefaultmidpunct}
{\mcitedefaultendpunct}{\mcitedefaultseppunct}\relax
\EndOfBibitem
\bibitem[{Ten-no} and Noga(2012){Ten-no}, and Noga]{VRG:ten-no:2012:WIRCMS}
{Ten-no},~S.; Noga,~J. Explicitly Correlated Electronic Structure Theory from
  {{R12}}/{{F12}} Ans{\"a}tze. \emph{Wiley Interdiscip. Rev. Comput. Mol. Sci.}
  \textbf{2012}, \emph{2}, 114--125\relax
\mciteBstWouldAddEndPuncttrue
\mciteSetBstMidEndSepPunct{\mcitedefaultmidpunct}
{\mcitedefaultendpunct}{\mcitedefaultseppunct}\relax
\EndOfBibitem
\bibitem[{Ten-no}(2012)]{VRG:ten-no:2012:TCA}
{Ten-no},~S. Explicitly Correlated Wave Functions: Summary and Perspective.
  \emph{Theor. Chim. Acta} \textbf{2012}, \emph{131}, 1070\relax
\mciteBstWouldAddEndPuncttrue
\mciteSetBstMidEndSepPunct{\mcitedefaultmidpunct}
{\mcitedefaultendpunct}{\mcitedefaultseppunct}\relax
\EndOfBibitem
\bibitem[Pavo{\v s}evi{\'c} \latin{et~al.}(2016)Pavo{\v s}evi{\'c}, Pinski,
  Riplinger, Neese, and Valeev]{VRG:pavosevic:2016:JCP}
Pavo{\v s}evi{\'c},~F.; Pinski,~P.; Riplinger,~C.; Neese,~F.; Valeev,~E.~F.
  {{SparseMaps}}---{{A}} Systematic Infrastructure for Reduced-Scaling
  Electronic Structure Methods. {{IV}}. {{Linear-scaling}} Second-Order
  Explicitly Correlated Energy with Pair Natural Orbitals. \emph{J. Chem.
  Phys.} \textbf{2016}, \emph{144}, 144109\relax
\mciteBstWouldAddEndPuncttrue
\mciteSetBstMidEndSepPunct{\mcitedefaultmidpunct}
{\mcitedefaultendpunct}{\mcitedefaultseppunct}\relax
\EndOfBibitem
\bibitem[Ma and Werner(2018)Ma, and Werner]{VRG:ma:2018:JCTC}
Ma,~Q.; Werner,~H.-J. Scalable Electron Correlation Methods. 5. {{Parallel}}
  Perturbative Triples Correction for Explicitly Correlated Local Coupled
  Cluster with Pair Natural Orbitals. \emph{J. Chem. Theory Comput.}
  \textbf{2018}, \emph{14}, 198--215\relax
\mciteBstWouldAddEndPuncttrue
\mciteSetBstMidEndSepPunct{\mcitedefaultmidpunct}
{\mcitedefaultendpunct}{\mcitedefaultseppunct}\relax
\EndOfBibitem
\bibitem[Kumar \latin{et~al.}(2020)Kumar, Neese, and
  Valeev]{VRG:kumar:2020:JCP}
Kumar,~A.; Neese,~F.; Valeev,~E.~F. Explicitly Correlated Coupled Cluster
  Method for Accurate Treatment of Open-Shell Molecules with Hundreds of Atoms.
  \emph{J. Chem. Phys.} \textbf{2020}, \emph{153}, 094105\relax
\mciteBstWouldAddEndPuncttrue
\mciteSetBstMidEndSepPunct{\mcitedefaultmidpunct}
{\mcitedefaultendpunct}{\mcitedefaultseppunct}\relax
\EndOfBibitem
\bibitem[Wang \latin{et~al.}(2023)Wang, Guo, Neese, Valeev, Li, and
  Li]{VRG:wang:2023:JCTC}
Wang,~Y.; Guo,~Y.; Neese,~F.; Valeev,~E.~F.; Li,~W.; Li,~S.
  Cluster-in-{{Molecule Approach}} with {{Explicitly Correlated Methods}} for
  {{Large Molecules}}. \emph{J. Chem. Theory Comput.} \textbf{2023}, \emph{19},
  8076--8089\relax
\mciteBstWouldAddEndPuncttrue
\mciteSetBstMidEndSepPunct{\mcitedefaultmidpunct}
{\mcitedefaultendpunct}{\mcitedefaultseppunct}\relax
\EndOfBibitem
\bibitem[{Ten-no}(2004)]{VRG:ten-no:2004:JCP}
{Ten-no},~S. Explicitly Correlated Second Order Perturbation Theory:
  {{Introduction}} of a Rational Generator and Numerical Quadratures. \emph{J.
  Chem. Phys.} \textbf{2004}, \emph{121}, 117\relax
\mciteBstWouldAddEndPuncttrue
\mciteSetBstMidEndSepPunct{\mcitedefaultmidpunct}
{\mcitedefaultendpunct}{\mcitedefaultseppunct}\relax
\EndOfBibitem
\bibitem[Huang \latin{et~al.}(1998)Huang, Filippi, and
  Umrigar]{VRG:huang:1998:JCP}
Huang,~C.-J.; Filippi,~C.; Umrigar,~C.~J. Spin Contamination in Quantum {{Monte
  Carlo}} Wave Functions. \emph{J. Chem. Phys.} \textbf{1998}, \emph{108},
  8838--8847\relax
\mciteBstWouldAddEndPuncttrue
\mciteSetBstMidEndSepPunct{\mcitedefaultmidpunct}
{\mcitedefaultendpunct}{\mcitedefaultseppunct}\relax
\EndOfBibitem
\bibitem[Shiozaki and Werner(2010)Shiozaki, and Werner]{VRG:shiozaki:2010:JCPa}
Shiozaki,~T.; Werner,~H.-J. Communication: {{Second-order}} Multireference
  Perturbation Theory with Explicit Correlation: {{CASPT2-F12}}. \emph{J. Chem.
  Phys.} \textbf{2010}, \emph{133}, 141103\relax
\mciteBstWouldAddEndPuncttrue
\mciteSetBstMidEndSepPunct{\mcitedefaultmidpunct}
{\mcitedefaultendpunct}{\mcitedefaultseppunct}\relax
\EndOfBibitem
\bibitem[Shiozaki \latin{et~al.}(2011)Shiozaki, Knizia, and
  Werner]{VRG:shiozaki:2011:JCPa}
Shiozaki,~T.; Knizia,~G.; Werner,~H.-J. Explicitly Correlated Multireference
  Configuration Interaction: {{MRCI-F12}}. \emph{J. Chem. Phys.} \textbf{2011},
  \emph{134}, 034113\relax
\mciteBstWouldAddEndPuncttrue
\mciteSetBstMidEndSepPunct{\mcitedefaultmidpunct}
{\mcitedefaultendpunct}{\mcitedefaultseppunct}\relax
\EndOfBibitem
\bibitem[Fliegl \latin{et~al.}(2006)Fliegl, Hättig, and
  Klopper]{cmast:Fliegl:2006:JCP}
Fliegl,~H.; Hättig,~C.; Klopper,~W. Coupled-cluster response theory with
  linear-R12 corrections: The CC2-R12 model for excitation energies. \emph{The
  Journal of Chemical Physics} \textbf{2006}, \emph{124}\relax
\mciteBstWouldAddEndPuncttrue
\mciteSetBstMidEndSepPunct{\mcitedefaultmidpunct}
{\mcitedefaultendpunct}{\mcitedefaultseppunct}\relax
\EndOfBibitem
\bibitem[Klopper(1991)]{VRG:klopper:1991:CPL}
Klopper,~W. Orbital-Invariant Formulation of the {{MP2-R12}} Method.
  \emph{Chem. Phys. Lett.} \textbf{1991}, \emph{186}, 583--585\relax
\mciteBstWouldAddEndPuncttrue
\mciteSetBstMidEndSepPunct{\mcitedefaultmidpunct}
{\mcitedefaultendpunct}{\mcitedefaultseppunct}\relax
\EndOfBibitem
\bibitem[Neiss \latin{et~al.}(2006)Neiss, Hättig, and
  Klopper]{cmast:Neiss:2006:JCP}
Neiss,~C.; Hättig,~C.; Klopper,~W. Extensions of R12 corrections to CC2-R12
  for excited states. \emph{The Journal of Chemical Physics} \textbf{2006},
  \emph{125}\relax
\mciteBstWouldAddEndPuncttrue
\mciteSetBstMidEndSepPunct{\mcitedefaultmidpunct}
{\mcitedefaultendpunct}{\mcitedefaultseppunct}\relax
\EndOfBibitem
\bibitem[Neiss and H{\"a}ttig(2007)Neiss, and H{\"a}ttig]{VRG:neiss:2007:JCP}
Neiss,~C.; H{\"a}ttig,~C. Frequency-Dependent Nonlinear Optical Properties with
  Explicitly Correlated Coupled-Cluster Response Theory Using the
  {{CCSD}}({{R12}}) Model. \emph{J. Chem. Phys.} \textbf{2007}, \emph{126},
  154101--154101--11\relax
\mciteBstWouldAddEndPuncttrue
\mciteSetBstMidEndSepPunct{\mcitedefaultmidpunct}
{\mcitedefaultendpunct}{\mcitedefaultseppunct}\relax
\EndOfBibitem
\bibitem[Köhn(2009)]{cmast:Kohn:2009:JCP}
Köhn,~A. A modified ansatz for explicitly correlated coupled-cluster wave
  functions that is suitable for response theory. \emph{The Journal of Chemical
  Physics} \textbf{2009}, \emph{130}\relax
\mciteBstWouldAddEndPuncttrue
\mciteSetBstMidEndSepPunct{\mcitedefaultmidpunct}
{\mcitedefaultendpunct}{\mcitedefaultseppunct}\relax
\EndOfBibitem
\bibitem[Höfener \latin{et~al.}(2019)Höfener, Schieschke, Klopper, and
  Köhn]{cmast:Kohn:2019:JCP}
Höfener,~S.; Schieschke,~N.; Klopper,~W.; Köhn,~A. The extended
  explicitly-correlated second-order approximate coupled-cluster singles and
  doubles ansatz suitable for response theory. \emph{The Journal of Chemical
  Physics} \textbf{2019}, \emph{150}\relax
\mciteBstWouldAddEndPuncttrue
\mciteSetBstMidEndSepPunct{\mcitedefaultmidpunct}
{\mcitedefaultendpunct}{\mcitedefaultseppunct}\relax
\EndOfBibitem
\bibitem[Bokhan and Ten-no(2010)Bokhan, and Ten-no]{cmast:Ten-no:2010:JCP}
Bokhan,~D.; Ten-no,~S. Explicitly correlated equation-of-motion coupled-cluster
  methods for excited and electron-attached states. \emph{The Journal of
  Chemical Physics} \textbf{2010}, \emph{133}\relax
\mciteBstWouldAddEndPuncttrue
\mciteSetBstMidEndSepPunct{\mcitedefaultmidpunct}
{\mcitedefaultendpunct}{\mcitedefaultseppunct}\relax
\EndOfBibitem
\bibitem[Bokhan \latin{et~al.}(2015)Bokhan, Trubnikov, and
  Bartlett]{cmast:Bartlett:2015:JCP}
Bokhan,~D.; Trubnikov,~D.~N.; Bartlett,~R.~J. Explicitly correlated similarity
  transformed equation-of-motion coupled-cluster method. \emph{The Journal of
  Chemical Physics} \textbf{2015}, \emph{143}\relax
\mciteBstWouldAddEndPuncttrue
\mciteSetBstMidEndSepPunct{\mcitedefaultmidpunct}
{\mcitedefaultendpunct}{\mcitedefaultseppunct}\relax
\EndOfBibitem
\bibitem[Pavo{\v s}evi{\'c} \latin{et~al.}(2017)Pavo{\v s}evi{\'c}, Peng,
  Ortiz, and Valeev]{VRG:pavosevic:2017:JCP}
Pavo{\v s}evi{\'c},~F.; Peng,~C.; Ortiz,~J.~V.; Valeev,~E.~F. Explicitly
  Correlated Formalism for Second-Order Single-Particle {{Green}}'s Function.
  \emph{J. Chem. Phys.} \textbf{2017}, \emph{147}, 121101\relax
\mciteBstWouldAddEndPuncttrue
\mciteSetBstMidEndSepPunct{\mcitedefaultmidpunct}
{\mcitedefaultendpunct}{\mcitedefaultseppunct}\relax
\EndOfBibitem
\bibitem[Teke \latin{et~al.}(2019)Teke, Pavo{\v s}evi{\'c}, Peng, and
  Valeev]{VRG:teke:2019:JCP}
Teke,~N.~K.; Pavo{\v s}evi{\'c},~F.; Peng,~C.; Valeev,~E.~F. Explicitly
  Correlated Renormalized Second-Order {{Green}}'s Function for Accurate
  Ionization Potentials of Closed-Shell Molecules. \emph{J. Chem. Phys.}
  \textbf{2019}, \emph{150}, 214103\relax
\mciteBstWouldAddEndPuncttrue
\mciteSetBstMidEndSepPunct{\mcitedefaultmidpunct}
{\mcitedefaultendpunct}{\mcitedefaultseppunct}\relax
\EndOfBibitem
\bibitem[Werner \latin{et~al.}(2011)Werner, Knizia, and
  Manby]{cmast:Manby:2011:MolPhys}
Werner,~H.-J.; Knizia,~G.; Manby,~F.~R. Explicitly correlated coupled cluster
  methods with pair-specific geminals. \emph{Molecular Physics} \textbf{2011},
  \emph{109}, 407–417\relax
\mciteBstWouldAddEndPuncttrue
\mciteSetBstMidEndSepPunct{\mcitedefaultmidpunct}
{\mcitedefaultendpunct}{\mcitedefaultseppunct}\relax
\EndOfBibitem
\bibitem[Yanai and Shiozaki(2012)Yanai, and Shiozaki]{VRG:yanai:2012:JCP}
Yanai,~T.; Shiozaki,~T. Canonical Transcorrelated Theory with Projected
  {{Slater-type}} Geminals. \emph{J. Chem. Phys.} \textbf{2012}, \emph{136},
  084107\relax
\mciteBstWouldAddEndPuncttrue
\mciteSetBstMidEndSepPunct{\mcitedefaultmidpunct}
{\mcitedefaultendpunct}{\mcitedefaultseppunct}\relax
\EndOfBibitem
\bibitem[Motta \latin{et~al.}(2020)Motta, Gujarati, Rice, Kumar, Masteran,
  Latone, Lee, Valeev, and Takeshita]{VRG:motta:2020:PCCP}
Motta,~M.; Gujarati,~T.~P.; Rice,~J.~E.; Kumar,~A.; Masteran,~C.;
  Latone,~J.~A.; Lee,~E.; Valeev,~E.~F.; Takeshita,~T.~Y. Quantum Simulation of
  Electronic Structure with a Transcorrelated {{Hamiltonian}}: Improved
  Accuracy with a Smaller Footprint on the Quantum Computer. \emph{Phys. Chem.
  Chem. Phys.} \textbf{2020}, \emph{22}, 24270--24281\relax
\mciteBstWouldAddEndPuncttrue
\mciteSetBstMidEndSepPunct{\mcitedefaultmidpunct}
{\mcitedefaultendpunct}{\mcitedefaultseppunct}\relax
\EndOfBibitem
\bibitem[Kersten \latin{et~al.}(2016)Kersten, Booth, and
  Alavi]{VRG:kersten:2016:JCP}
Kersten,~J. A.~F.; Booth,~G.~H.; Alavi,~A. Assessment of Multireference
  Approaches to Explicitly Correlated Full Configuration Interaction Quantum
  {{Monte Carlo}}. \emph{J. Chem. Phys.} \textbf{2016}, \emph{145},
  054117\relax
\mciteBstWouldAddEndPuncttrue
\mciteSetBstMidEndSepPunct{\mcitedefaultmidpunct}
{\mcitedefaultendpunct}{\mcitedefaultseppunct}\relax
\EndOfBibitem
\bibitem[Masteran \latin{et~al.}(2023)Masteran, Kumar, Teke, Gaudel, Yanai, and
  Valeev]{cmast:Masteran:2023:JCP}
Masteran,~C.; Kumar,~A.; Teke,~N.; Gaudel,~B.; Yanai,~T.; Valeev,~E.~F. Comment
  on “Canonical transcorrelated theory with projected Slater-type geminals”
  [J. Chem. Phys. 136, 084107 (2012)]. \emph{The Journal of Chemical Physics}
  \textbf{2023}, \emph{158}\relax
\mciteBstWouldAddEndPuncttrue
\mciteSetBstMidEndSepPunct{\mcitedefaultmidpunct}
{\mcitedefaultendpunct}{\mcitedefaultseppunct}\relax
\EndOfBibitem
\bibitem[Handy(1969)]{VRG:handy:1969:JCP}
Handy,~N.~C. Energies and Expectation Values for Be by the Transcorrelated
  Method. \emph{J. Chem. Phys.} \textbf{1969}, \emph{51}, 3205--3212\relax
\mciteBstWouldAddEndPuncttrue
\mciteSetBstMidEndSepPunct{\mcitedefaultmidpunct}
{\mcitedefaultendpunct}{\mcitedefaultseppunct}\relax
\EndOfBibitem
\bibitem[Handy(1971)]{VRG:handy:1971:MP}
Handy,~N.~C. On the Minimization of the Variance of the Transcorrelated
  Hamiltonian. \emph{Mol. Phys.} \textbf{1971}, \emph{21}, 817--828\relax
\mciteBstWouldAddEndPuncttrue
\mciteSetBstMidEndSepPunct{\mcitedefaultmidpunct}
{\mcitedefaultendpunct}{\mcitedefaultseppunct}\relax
\EndOfBibitem
\bibitem[Schraivogel \latin{et~al.}(2021)Schraivogel, Cohen, Alavi, and
  Kats]{cmast:Schraivogel:2021}
Schraivogel,~T.; Cohen,~A.~J.; Alavi,~A.; Kats,~D. Transcorrelated coupled
  cluster methods. \emph{The Journal of Chemical Physics} \textbf{2021},
  \emph{155}\relax
\mciteBstWouldAddEndPuncttrue
\mciteSetBstMidEndSepPunct{\mcitedefaultmidpunct}
{\mcitedefaultendpunct}{\mcitedefaultseppunct}\relax
\EndOfBibitem
\bibitem[Baiardi \latin{et~al.}(2022)Baiardi, Lesiuk, and
  Reiher]{VRG:baiardi:2022:JCTC}
Baiardi,~A.; Lesiuk,~M.; Reiher,~M. Explicitly {{Correlated Electronic
  Structure Calculations}} with {{Transcorrelated Matrix Product Operators}}.
  \emph{J. Chem. Theory Comput.} \textbf{2022}, \emph{18}, 4203--4217\relax
\mciteBstWouldAddEndPuncttrue
\mciteSetBstMidEndSepPunct{\mcitedefaultmidpunct}
{\mcitedefaultendpunct}{\mcitedefaultseppunct}\relax
\EndOfBibitem
\bibitem[Kats \latin{et~al.}(2024)Kats, Christlmaier, Schraivogel, and
  Alavi]{VRG:kats:2024:FD}
Kats,~D.; Christlmaier,~E. M.~C.; Schraivogel,~T.; Alavi,~A. Orbital
  Optimisation in {{xTC}} Transcorrelated Methods. \emph{Faraday Discuss.}
  \textbf{2024}, \emph{254}, 382--401\relax
\mciteBstWouldAddEndPuncttrue
\mciteSetBstMidEndSepPunct{\mcitedefaultmidpunct}
{\mcitedefaultendpunct}{\mcitedefaultseppunct}\relax
\EndOfBibitem
\bibitem[Szenes \latin{et~al.}(2024)Szenes, M{\"o}rchen, Fischill, and
  Reiher]{VRG:szenes:2024:FD}
Szenes,~K.; M{\"o}rchen,~M.; Fischill,~P.; Reiher,~M. Striking the Right
  Balance of Encoding Electron Correlation in the {{Hamiltonian}} and the
  Wavefunction Ansatz. \emph{Faraday Discuss.} \textbf{2024}, \emph{254},
  359--381\relax
\mciteBstWouldAddEndPuncttrue
\mciteSetBstMidEndSepPunct{\mcitedefaultmidpunct}
{\mcitedefaultendpunct}{\mcitedefaultseppunct}\relax
\EndOfBibitem
\bibitem[Lee and Thom(2023)Lee, and Thom]{VRG:lee:2023:JCTC}
Lee,~N.; Thom,~A. J.~W. Studies on the {{Transcorrelated Method}}. \emph{J.
  Chem. Theory Comput.} \textbf{2023}, \emph{19}, 5743--5759\relax
\mciteBstWouldAddEndPuncttrue
\mciteSetBstMidEndSepPunct{\mcitedefaultmidpunct}
{\mcitedefaultendpunct}{\mcitedefaultseppunct}\relax
\EndOfBibitem
\bibitem[{Ten-no}(2023)]{VRG:ten-no:2023:JCP}
{Ten-no},~S.~L. Nonunitary Projective Transcorrelation Theory Inspired by the
  {{F12}} Ansatz. \emph{J. Chem. Phys.} \textbf{2023}, \emph{159}, 171103\relax
\mciteBstWouldAddEndPuncttrue
\mciteSetBstMidEndSepPunct{\mcitedefaultmidpunct}
{\mcitedefaultendpunct}{\mcitedefaultseppunct}\relax
\EndOfBibitem
\bibitem[Hino \latin{et~al.}(2001)Hino, Tanimura, and Ten-no]{cmast:Hino:2001}
Hino,~O.; Tanimura,~Y.; Ten-no,~S. Biorthogonal approach for explicitly
  correlated calculations using the transcorrelated Hamiltonian. \emph{The
  Journal of Chemical Physics} \textbf{2001}, \emph{115}, 7865–7871\relax
\mciteBstWouldAddEndPuncttrue
\mciteSetBstMidEndSepPunct{\mcitedefaultmidpunct}
{\mcitedefaultendpunct}{\mcitedefaultseppunct}\relax
\EndOfBibitem
\bibitem[Hino \latin{et~al.}(2002)Hino, Tanimura, and Ten-no]{cmast:Hino:2002}
Hino,~O.; Tanimura,~Y.; Ten-no,~S. Application of the transcorrelated
  Hamiltonian to the linearized coupled cluster singles and doubles model.
  \emph{Chemical Physics Letters} \textbf{2002}, \emph{353}, 317–323\relax
\mciteBstWouldAddEndPuncttrue
\mciteSetBstMidEndSepPunct{\mcitedefaultmidpunct}
{\mcitedefaultendpunct}{\mcitedefaultseppunct}\relax
\EndOfBibitem
\bibitem[Cohen \latin{et~al.}(2019)Cohen, Luo, Guther, Dobrautz, Tew, and
  Alavi]{cmast:Cohen:2019}
Cohen,~A.~J.; Luo,~H.; Guther,~K.; Dobrautz,~W.; Tew,~D.~P.; Alavi,~A.
  Similarity transformation of the electronic Schr\"{o}dinger equation via
  Jastrow factorization. \emph{The Journal of Chemical Physics} \textbf{2019},
  \emph{151}\relax
\mciteBstWouldAddEndPuncttrue
\mciteSetBstMidEndSepPunct{\mcitedefaultmidpunct}
{\mcitedefaultendpunct}{\mcitedefaultseppunct}\relax
\EndOfBibitem
\bibitem[Nooijen and Bartlett(1998)Nooijen, and Bartlett]{VRG:nooijen:1998:JCP}
Nooijen,~M.; Bartlett,~R.~J. Elimination of {{Coulombic}} Infinities through
  Transformation of the {{Hamiltonian}}. \emph{J. Chem. Phys.} \textbf{1998},
  \emph{109}, 8232--8240\relax
\mciteBstWouldAddEndPuncttrue
\mciteSetBstMidEndSepPunct{\mcitedefaultmidpunct}
{\mcitedefaultendpunct}{\mcitedefaultseppunct}\relax
\EndOfBibitem
\bibitem[Valeev(2008)]{VRG:valeev:2008:PCCPP}
Valeev,~E.~F. Coupled-Cluster Methods with Perturbative Inclusion of Explicitly
  Correlated Terms: A Preliminary Investigation. \emph{Phys. Chem. Chem. Phys.}
  \textbf{2008}, \emph{10}, 106--113\relax
\mciteBstWouldAddEndPuncttrue
\mciteSetBstMidEndSepPunct{\mcitedefaultmidpunct}
{\mcitedefaultendpunct}{\mcitedefaultseppunct}\relax
\EndOfBibitem
\bibitem[Adler \latin{et~al.}(2007)Adler, Knizia, and
  Werner]{VRG:adler:2007:JCP}
Adler,~T.~B.; Knizia,~G.; Werner,~H.-J. A Simple and Efficient
  {{CCSD}}({{T}})-{{F12}} Approximation. \emph{J. Chem. Phys.} \textbf{2007},
  \emph{127}, 221106\relax
\mciteBstWouldAddEndPuncttrue
\mciteSetBstMidEndSepPunct{\mcitedefaultmidpunct}
{\mcitedefaultendpunct}{\mcitedefaultseppunct}\relax
\EndOfBibitem
\bibitem[H{\"a}ttig \latin{et~al.}(2010)H{\"a}ttig, Tew, and
  K{\"o}hn]{VRG:hattig:2010:JCP}
H{\"a}ttig,~C.; Tew,~D.~P.; K{\"o}hn,~A. Communications: {{Accurate}} and
  Efficient Approximations to Explicitly Correlated Coupled-Cluster Singles and
  Doubles, {{CCSD-F12}}. \emph{J. Chem. Phys.} \textbf{2010}, \emph{132},
  231102--231102--4\relax
\mciteBstWouldAddEndPuncttrue
\mciteSetBstMidEndSepPunct{\mcitedefaultmidpunct}
{\mcitedefaultendpunct}{\mcitedefaultseppunct}\relax
\EndOfBibitem
\bibitem[Zhang and Valeev(2012)Zhang, and Valeev]{VRG:zhang:2012:JCTC}
Zhang,~J.; Valeev,~E.~F. Prediction of Reaction Barriers and Thermochemical
  Properties with Explicitly Correlated Coupled-Cluster Methods: {{A}} Basis
  Set Assessment. \emph{J. Chem. Theory Comput.} \textbf{2012}, \emph{8},
  3175--3186\relax
\mciteBstWouldAddEndPuncttrue
\mciteSetBstMidEndSepPunct{\mcitedefaultmidpunct}
{\mcitedefaultendpunct}{\mcitedefaultseppunct}\relax
\EndOfBibitem
\bibitem[Valeev and Janssen(2004)Valeev, and Janssen]{VRG:valeev:2004:JCP}
Valeev,~E.~F.; Janssen,~C.~L. Second-Order {{M{\o}ller-Plesset}} Theory with
  Linear {{R12}} Terms ({{MP2-R12}}) Revisited: Auxiliary Basis Set Method and
  Massively Parallel Implementation. \emph{J. Chem. Phys.} \textbf{2004},
  \emph{121}, 1214--1227\relax
\mciteBstWouldAddEndPuncttrue
\mciteSetBstMidEndSepPunct{\mcitedefaultmidpunct}
{\mcitedefaultendpunct}{\mcitedefaultseppunct}\relax
\EndOfBibitem
\bibitem[Valeev(2004)]{VRG:valeev:2004:CPL}
Valeev,~E.~F. Improving on the Resolution of the Identity in Linear {{R12}} Ab
  Initio Theories. \emph{Chem. Phys. Lett.} \textbf{2004}, \emph{395},
  190--195\relax
\mciteBstWouldAddEndPuncttrue
\mciteSetBstMidEndSepPunct{\mcitedefaultmidpunct}
{\mcitedefaultendpunct}{\mcitedefaultseppunct}\relax
\EndOfBibitem
\bibitem[Werner and Manby(2006)Werner, and Manby]{VRG:werner:2006:JCP}
Werner,~H.~J.; Manby,~F.~R. Explicitly Correlated Second-Order Perturbation
  Theory Using Density Fitting and Local Approximations. \emph{J. Chem. Phys.}
  \textbf{2006}, \relax
\mciteBstWouldAddEndPunctfalse
\mciteSetBstMidEndSepPunct{\mcitedefaultmidpunct}
{}{\mcitedefaultseppunct}\relax
\EndOfBibitem
\bibitem[Mukherjee(1997)]{cmast:Mukherjee:1997:CPL}
Mukherjee,~D. Normal ordering and a Wick-like reduction theorem for fermions
  with respect to a multi-determinantal reference state. \emph{Chemical Physics
  Letters} \textbf{1997}, \emph{274}, 561--566\relax
\mciteBstWouldAddEndPuncttrue
\mciteSetBstMidEndSepPunct{\mcitedefaultmidpunct}
{\mcitedefaultendpunct}{\mcitedefaultseppunct}\relax
\EndOfBibitem
\bibitem[Kutzelnigg and Mukherjee(1997)Kutzelnigg, and
  Mukherjee]{cmast:Kutzelnigg:1997:JCP}
Kutzelnigg,~W.; Mukherjee,~D. Normal order and extended Wick theorem for a
  multiconfiguration reference wave function. \emph{The Journal of Chemical
  Physics} \textbf{1997}, \emph{107}, 432--449\relax
\mciteBstWouldAddEndPuncttrue
\mciteSetBstMidEndSepPunct{\mcitedefaultmidpunct}
{\mcitedefaultendpunct}{\mcitedefaultseppunct}\relax
\EndOfBibitem
\bibitem[Kutzelnigg \latin{et~al.}(2010)Kutzelnigg, Shamasundar, and
  Mukherjee]{cmast:Kutzelnigg:2010}
Kutzelnigg,~W.; Shamasundar,~K.~R.; Mukherjee,~D. Spinfree formulation of
  reduced density matrices, density cumulants and generalised normal ordering.
  \emph{Molecular Physics} \textbf{2010}, \emph{108}, 433--451\relax
\mciteBstWouldAddEndPuncttrue
\mciteSetBstMidEndSepPunct{\mcitedefaultmidpunct}
{\mcitedefaultendpunct}{\mcitedefaultseppunct}\relax
\EndOfBibitem
\bibitem[Kumar \latin{et~al.}(2022)Kumar, Asthana, Masteran, Valeev, Zhang,
  Cincio, Tretiak, and Dub]{cmast:Kumar:2022}
Kumar,~A.; Asthana,~A.; Masteran,~C.; Valeev,~E.~F.; Zhang,~Y.; Cincio,~L.;
  Tretiak,~S.; Dub,~P.~A. Quantum Simulation of Molecular Electronic States
  with a Transcorrelated Hamiltonian: Higher Accuracy with Fewer Qubits.
  \emph{Journal of Chemical Theory and Computation} \textbf{2022}, \emph{18},
  5312–5324\relax
\mciteBstWouldAddEndPuncttrue
\mciteSetBstMidEndSepPunct{\mcitedefaultmidpunct}
{\mcitedefaultendpunct}{\mcitedefaultseppunct}\relax
\EndOfBibitem
\bibitem[Ked{\v z}uch \latin{et~al.}(2005)Ked{\v z}uch, Milko, and
  Noga]{VRG:kedzuch:2005:IJQC}
Ked{\v z}uch,~S.; Milko,~M.; Noga,~J. Alternative Formulation of the Matrix
  Elements in {{MP2-R12}} Theory. \emph{Int. J. Quantum Chem.} \textbf{2005},
  \emph{105}, 929--936\relax
\mciteBstWouldAddEndPuncttrue
\mciteSetBstMidEndSepPunct{\mcitedefaultmidpunct}
{\mcitedefaultendpunct}{\mcitedefaultseppunct}\relax
\EndOfBibitem
\bibitem[Lu \latin{et~al.}(2004)Lu, Wang, Schmidt, Bytautas, Ho, and
  Ruedenberg]{VRG:lu:2004:JCP}
Lu,~W.~C.; Wang,~C.~Z.; Schmidt,~M.~W.; Bytautas,~L.; Ho,~K.~M.; Ruedenberg,~K.
  Molecule Intrinsic Minimal Basis Sets. {{I}}. {{Exact}} Resolution of
  {\emph{Ab Initio}} Optimized Molecular Orbitals in Terms of Deformed Atomic
  Minimal-Basis Orbitals. \emph{J. Chem. Phys.} \textbf{2004}, \emph{120},
  2629--2637\relax
\mciteBstWouldAddEndPuncttrue
\mciteSetBstMidEndSepPunct{\mcitedefaultmidpunct}
{\mcitedefaultendpunct}{\mcitedefaultseppunct}\relax
\EndOfBibitem
\bibitem[Schmidt \latin{et~al.}(2015)Schmidt, Hull, and
  Windus]{cmast:Windus:2015:JPCA}
Schmidt,~M.~W.; Hull,~E.~A.; Windus,~T.~L. Valence virtual orbitals: An
  unambiguous ab initio quantification of the LUMO concept. \emph{The Journal
  of Physical Chemistry A} \textbf{2015}, \emph{119}, 10408–10427\relax
\mciteBstWouldAddEndPuncttrue
\mciteSetBstMidEndSepPunct{\mcitedefaultmidpunct}
{\mcitedefaultendpunct}{\mcitedefaultseppunct}\relax
\EndOfBibitem
\bibitem[Iwata(1981)]{VRG:iwata:1981:CPL}
Iwata,~S. Valence Type Vacant Orbitals for Configuration Interaction
  Calculations. \emph{Chem. Phys. Lett.} \textbf{1981}, \emph{83},
  134--138\relax
\mciteBstWouldAddEndPuncttrue
\mciteSetBstMidEndSepPunct{\mcitedefaultmidpunct}
{\mcitedefaultendpunct}{\mcitedefaultseppunct}\relax
\EndOfBibitem
\bibitem[Sayfutyarova \latin{et~al.}(2017)Sayfutyarova, Sun, Chan, and
  Knizia]{VRG:sayfutyarova:2017:JCTC}
Sayfutyarova,~E.~R.; Sun,~Q.; Chan,~G. K.-L.; Knizia,~G. Automated
  {{Construction}} of {{Molecular Active Spaces}} from {{Atomic Valence
  Orbitals}}. \emph{J. Chem. Theory Comput.} \textbf{2017}, \emph{13},
  4063--4078\relax
\mciteBstWouldAddEndPuncttrue
\mciteSetBstMidEndSepPunct{\mcitedefaultmidpunct}
{\mcitedefaultendpunct}{\mcitedefaultseppunct}\relax
\EndOfBibitem
\bibitem[Peng \latin{et~al.}(2020)Peng, Lewis, Wang, Clement, Pierce, Rishi,
  Pavo{\v s}evi{\'c}, Slattery, Zhang, Teke, Kumar, Masteran, Asadchev, Calvin,
  and Valeev]{VRG:peng:2020:JCP}
Peng,~C.; Lewis,~C.~A.; Wang,~X.; Clement,~M.~C.; Pierce,~K.; Rishi,~V.;
  Pavo{\v s}evi{\'c},~F.; Slattery,~S.; Zhang,~J.; Teke,~N.; Kumar,~A.;
  Masteran,~C.; Asadchev,~A.; Calvin,~J.~A.; Valeev,~E.~F. Massively {{Parallel
  Quantum Chemistry}}: {{A}} High-Performance Research Platform for Electronic
  Structure. \emph{J. Chem. Phys.} \textbf{2020}, \emph{153}, 044120\relax
\mciteBstWouldAddEndPuncttrue
\mciteSetBstMidEndSepPunct{\mcitedefaultmidpunct}
{\mcitedefaultendpunct}{\mcitedefaultseppunct}\relax
\EndOfBibitem
\bibitem[{Ten-no}(2004)]{VRG:ten-no:2004:CPL}
{Ten-no},~S. Initiation of Explicitly Correlated {{Slater-type}} Geminal
  Theory. \emph{Chem. Phys. Lett.} \textbf{2004}, \emph{398}, 56--61\relax
\mciteBstWouldAddEndPuncttrue
\mciteSetBstMidEndSepPunct{\mcitedefaultmidpunct}
{\mcitedefaultendpunct}{\mcitedefaultseppunct}\relax
\EndOfBibitem
\bibitem[May and Manby(2004)May, and Manby]{VRG:may:2004:JCP}
May,~A.~J.; Manby,~F.~R. An Explicitly Correlated Second Order
  {{M{\o}ller-Plesset}} Theory Using a Frozen {{Gaussian}} Geminal. \emph{J.
  Chem. Phys.} \textbf{2004}, \emph{121}, 4479--4485\relax
\mciteBstWouldAddEndPuncttrue
\mciteSetBstMidEndSepPunct{\mcitedefaultmidpunct}
{\mcitedefaultendpunct}{\mcitedefaultseppunct}\relax
\EndOfBibitem
\bibitem[Tew and Klopper(2005)Tew, and Klopper]{VRG:tew:2005:JCP}
Tew,~D.~P.; Klopper,~W. New Correlation Factors for Explicitly Correlated
  Electronic Wave Functions. \emph{J. Chem. Phys.} \textbf{2005}, \emph{123},
  074101\relax
\mciteBstWouldAddEndPuncttrue
\mciteSetBstMidEndSepPunct{\mcitedefaultmidpunct}
{\mcitedefaultendpunct}{\mcitedefaultseppunct}\relax
\EndOfBibitem
\bibitem[Dunning(1989)]{cmast:Dunning:1989}
Dunning,~T.~H. Gaussian basis sets for use in correlated molecular
  calculations. I. The atoms boron through neon and hydrogen. \emph{The Journal
  of Chemical Physics} \textbf{1989}, \emph{90}, 1007–1023\relax
\mciteBstWouldAddEndPuncttrue
\mciteSetBstMidEndSepPunct{\mcitedefaultmidpunct}
{\mcitedefaultendpunct}{\mcitedefaultseppunct}\relax
\EndOfBibitem
\bibitem[Kendall \latin{et~al.}(1992)Kendall, Dunning, and
  Harrison]{cmast:Kendall:1992}
Kendall,~R.~A.; Dunning,~T.~H.; Harrison,~R.~J. Electron affinities of the
  first-row atoms revisited. Systematic basis sets and wave functions.
  \emph{The Journal of Chemical Physics} \textbf{1992}, \emph{96},
  6796–6806\relax
\mciteBstWouldAddEndPuncttrue
\mciteSetBstMidEndSepPunct{\mcitedefaultmidpunct}
{\mcitedefaultendpunct}{\mcitedefaultseppunct}\relax
\EndOfBibitem
\bibitem[Woon and Dunning(1994)Woon, and Dunning]{cmast:Woon:1994}
Woon,~D.~E.; Dunning,~T.~H. Gaussian basis sets for use in correlated molecular
  calculations. IV. Calculation of static electrical response properties.
  \emph{The Journal of Chemical Physics} \textbf{1994}, \emph{100},
  2975–2988\relax
\mciteBstWouldAddEndPuncttrue
\mciteSetBstMidEndSepPunct{\mcitedefaultmidpunct}
{\mcitedefaultendpunct}{\mcitedefaultseppunct}\relax
\EndOfBibitem
\bibitem[Peterson \latin{et~al.}(2008)Peterson, Adler, and
  Werner]{cmast:Peterson:2008}
Peterson,~K.~A.; Adler,~T.~B.; Werner,~H.-J. Systematically convergent basis
  sets for explicitly correlated wavefunctions: The atoms H, he, B–ne, and
  al–ar. \emph{The Journal of Chemical Physics} \textbf{2008},
  \emph{128}\relax
\mciteBstWouldAddEndPuncttrue
\mciteSetBstMidEndSepPunct{\mcitedefaultmidpunct}
{\mcitedefaultendpunct}{\mcitedefaultseppunct}\relax
\EndOfBibitem
\bibitem[Yousaf and Peterson(2009)Yousaf, and Peterson]{cmast:Peterson:2009}
Yousaf,~K.~E.; Peterson,~K.~A. Optimized complementary auxiliary basis sets for
  explicitly correlated methods: Aug-cc-pvnz orbital basis sets. \emph{Chemical
  Physics Letters} \textbf{2009}, \emph{476}, 303–307\relax
\mciteBstWouldAddEndPuncttrue
\mciteSetBstMidEndSepPunct{\mcitedefaultmidpunct}
{\mcitedefaultendpunct}{\mcitedefaultseppunct}\relax
\EndOfBibitem
\bibitem[Weigend \latin{et~al.}(2002)Weigend, K\"{o}hn, and
  H\"{a}ttig]{cmast:Weigend:2002}
Weigend,~F.; K\"{o}hn,~A.; H\"{a}ttig,~C. Efficient use of the correlation
  consistent basis sets in resolution of the identity MP2 calculations.
  \emph{The Journal of Chemical Physics} \textbf{2002}, \emph{116},
  3175–3183\relax
\mciteBstWouldAddEndPuncttrue
\mciteSetBstMidEndSepPunct{\mcitedefaultmidpunct}
{\mcitedefaultendpunct}{\mcitedefaultseppunct}\relax
\EndOfBibitem
\bibitem[Andzelm \latin{et~al.}(1984)Andzelm, Huzinaga, Klobukowski,
  {Radzio-Andzelm}, Sakai, and Tatewaki]{VRG:andzelm:1984:PSDV1GBSfMC}
Andzelm,~J.; Huzinaga,~S.; Klobukowski,~M.; {Radzio-Andzelm},~E.; Sakai,~Y.;
  Tatewaki,~H. In \emph{Physical {{Sciences Data}}, {{Volume}} 16: {{Gaussian
  Basis Sets}} for {{Molecular Calculations}}}; Huzinaga,~S., Ed.; Elsevier,
  1984; Vol.~16; pp 27--426\relax
\mciteBstWouldAddEndPuncttrue
\mciteSetBstMidEndSepPunct{\mcitedefaultmidpunct}
{\mcitedefaultendpunct}{\mcitedefaultseppunct}\relax
\EndOfBibitem
\bibitem[{van Duijneveldt}(1971)]{VRG:vanduijneveldt:1971:}
{van Duijneveldt},~F.~B. \emph{Gaussian Basis Sets for the Atoms {{H-Ne}} for
  Use in Molecular Calculations}; 1971\relax
\mciteBstWouldAddEndPuncttrue
\mciteSetBstMidEndSepPunct{\mcitedefaultmidpunct}
{\mcitedefaultendpunct}{\mcitedefaultseppunct}\relax
\EndOfBibitem
\bibitem[Nakatsuji(1979)]{VRG:nakatsuji:1979:}
Nakatsuji,~H. Cluster Expansion of the Wavefunction. {{Electron}} Correlations
  in Ground and Excited States by {{SAC}} (Symmetry-Adapted-Cluster) and {{SAC
  CI}} Theories. 1979;
  \url{http://www.sciencedirect.com/science/article/pii/0009261479851726}\relax
\mciteBstWouldAddEndPuncttrue
\mciteSetBstMidEndSepPunct{\mcitedefaultmidpunct}
{\mcitedefaultendpunct}{\mcitedefaultseppunct}\relax
\EndOfBibitem
\bibitem[Stanton and Bartlett(1993)Stanton, and Bartlett]{cmast:Stanton1993}
Stanton,~J.~F.; Bartlett,~R.~J. The equation of motion coupled-cluster method.
  A systematic biorthogonal approach to molecular excitation energies,
  transition probabilities, and excited state properties. \emph{The Journal of
  Chemical Physics} \textbf{1993}, \emph{98}, 7029–7039\relax
\mciteBstWouldAddEndPuncttrue
\mciteSetBstMidEndSepPunct{\mcitedefaultmidpunct}
{\mcitedefaultendpunct}{\mcitedefaultseppunct}\relax
\EndOfBibitem
\bibitem[Peng \latin{et~al.}(2018)Peng, Clement, and
  Valeev]{VRG:peng:2018:JCTC}
Peng,~C.; Clement,~M.~C.; Valeev,~E.~F. State-Averaged Pair Natural Orbitals
  for Excited States: {{A}} Route toward Efficient Equation of Motion
  Coupled-Cluster. \emph{J. Chem. Theory Comput.} \textbf{2018}, \emph{14},
  5597--5607\relax
\mciteBstWouldAddEndPuncttrue
\mciteSetBstMidEndSepPunct{\mcitedefaultmidpunct}
{\mcitedefaultendpunct}{\mcitedefaultseppunct}\relax
\EndOfBibitem
\bibitem[Yang and H{\"a}ttig(2009)Yang, and H{\"a}ttig]{cmast:Yang:2009}
Yang,~J.; H{\"a}ttig,~C. Structures and harmonic vibrational frequencies for
  excited states of diatomic molecules with CCSD(R12) and CCSD(F12) models.
  \emph{Journal of Chemical Physics} \textbf{2009}, \emph{130}, 124101\relax
\mciteBstWouldAddEndPuncttrue
\mciteSetBstMidEndSepPunct{\mcitedefaultmidpunct}
{\mcitedefaultendpunct}{\mcitedefaultseppunct}\relax
\EndOfBibitem
\bibitem[Schreiber \latin{et~al.}(2008)Schreiber, Silva-Junior, Sauer, and
  Thiel]{cmast:Schreiber:2008}
Schreiber,~M.; Silva-Junior,~M.~R.; Sauer,~S. P.~A.; Thiel,~W. Benchmarks for
  electronically excited states: CASPT2, CC2, CCSD, and CC3. \emph{The Journal
  of Chemical Physics} \textbf{2008}, \emph{128}\relax
\mciteBstWouldAddEndPuncttrue
\mciteSetBstMidEndSepPunct{\mcitedefaultmidpunct}
{\mcitedefaultendpunct}{\mcitedefaultseppunct}\relax
\EndOfBibitem
\bibitem[Tajti \latin{et~al.}(2004)Tajti, Szalay, Császár, Kállay, Gauss,
  Valeev, Flowers, Vázquez, and Stanton]{cmast:Tajti2004}
Tajti,~A.; Szalay,~P.~G.; Császár,~A.~G.; Kállay,~M.; Gauss,~J.;
  Valeev,~E.~F.; Flowers,~B.~A.; Vázquez,~J.; Stanton,~J.~F. HEAT: High
  accuracy extrapolated ab initio thermochemistry. \emph{The Journal of
  Chemical Physics} \textbf{2004}, \emph{121}, 11599–11613\relax
\mciteBstWouldAddEndPuncttrue
\mciteSetBstMidEndSepPunct{\mcitedefaultmidpunct}
{\mcitedefaultendpunct}{\mcitedefaultseppunct}\relax
\EndOfBibitem
\bibitem[Knizia and Werner(2008)Knizia, and Werner]{VRG:knizia:2008:JCP}
Knizia,~G.; Werner,~H.-J. Explicitly Correlated {{RMP2}} for High-Spin
  Open-Shell Reference States. \emph{J. Chem. Phys.} \textbf{2008}, \emph{128},
  154103\relax
\mciteBstWouldAddEndPuncttrue
\mciteSetBstMidEndSepPunct{\mcitedefaultmidpunct}
{\mcitedefaultendpunct}{\mcitedefaultseppunct}\relax
\EndOfBibitem
\bibitem[Wolinski and Pulay(2003)Wolinski, and Pulay]{VRG:wolinski:2003:JCP}
Wolinski,~K.; Pulay,~P. Second-Order {{M{\o}ller}}--{{Plesset}} Calculations
  with Dual Basis Sets. \textbf{2003}, \emph{118}, 9497--9503\relax
\mciteBstWouldAddEndPuncttrue
\mciteSetBstMidEndSepPunct{\mcitedefaultmidpunct}
{\mcitedefaultendpunct}{\mcitedefaultseppunct}\relax
\EndOfBibitem
\bibitem[Shiozaki \latin{et~al.}(2008)Shiozaki, Kamiya, Hirata, and
  Valeev]{VRG:shiozaki:2008:JCP}
Shiozaki,~T.; Kamiya,~M.; Hirata,~S.; Valeev,~E.~F. Explicitly Correlated
  Coupled-Cluster Singles and Doubles Method Based on Complete Diagrammatic
  Equations. \emph{J. Chem. Phys.} \textbf{2008}, \emph{129}, 071101\relax
\mciteBstWouldAddEndPuncttrue
\mciteSetBstMidEndSepPunct{\mcitedefaultmidpunct}
{\mcitedefaultendpunct}{\mcitedefaultseppunct}\relax
\EndOfBibitem
\bibitem[K{\"o}hn \latin{et~al.}(2008)K{\"o}hn, Richings, and
  Tew]{VRG:kohn:2008:JCP}
K{\"o}hn,~A.; Richings,~G.~W.; Tew,~D.~P. Implementation of the Full Explicitly
  Correlated Coupled-Cluster Singles and Doubles Model {{CCSD-F12}} with
  Optimally Reduced Auxiliary Basis Dependence. \emph{J. Chem. Phys.}
  \textbf{2008}, \emph{129}, 201103\relax
\mciteBstWouldAddEndPuncttrue
\mciteSetBstMidEndSepPunct{\mcitedefaultmidpunct}
{\mcitedefaultendpunct}{\mcitedefaultseppunct}\relax
\EndOfBibitem
\bibitem[Powell and Valeev(2025)Powell, and Valeev]{cmast:Powell:2025}
Powell,~S.~R.; Valeev,~E.~F. Slimmer Geminals For Accurate F12 Electronic
  Structure Models. 2025; \url{https://arxiv.org/abs/2506.10780}\relax
\mciteBstWouldAddEndPuncttrue
\mciteSetBstMidEndSepPunct{\mcitedefaultmidpunct}
{\mcitedefaultendpunct}{\mcitedefaultseppunct}\relax
\EndOfBibitem
\bibitem[Klopper \latin{et~al.}(2006)Klopper, Manby, {Ten-no}, and
  Valeev]{VRG:klopper:2006:IRPC}
Klopper,~W.; Manby,~F.~R.; {Ten-no},~S.; Valeev,~E.~F. R12 Methods in
  Explicitly Correlated Molecular Electronic Structure Theory. \emph{Int. Rev.
  Phys. Chem.} \textbf{2006}, \emph{25}, 427--468\relax
\mciteBstWouldAddEndPuncttrue
\mciteSetBstMidEndSepPunct{\mcitedefaultmidpunct}
{\mcitedefaultendpunct}{\mcitedefaultseppunct}\relax
\EndOfBibitem
\end{mcitethebibliography}
\end{document}